\documentclass{jfm}

\usepackage{xspace}
\usepackage{algorithm}
\usepackage{algorithmic}
\usepackage{diagbox}
\usepackage{subcaption}
\usepackage[textsize=tiny,disable]{todonotes} 

\newcommand{\vesnet}{VesNet\xspace}
\newcommand{\ivnet}{IVNet\xspace}

\newcommand{\uu}{\mathbf{u}}

\newcommand{\XX}{\mathbf{X}}

\newcommand{\xx}{\mathbf{x}}
\newcommand{\nn}{\mathbf{n}}

\newcommand{\rr}{\mathbf{r}}

\newcommand{\ubar}{\overline{\mathbf{u}}}

\newcommand{\ff}{\mathbf{f}}

\newcommand{\GG}{\mathbf{G}}

\newcommand{\calB}{\mathcal{B}}
\newcommand{\calL}{\mathcal{L}}

\newcommand{\calT}{\mathcal{T}}
\newcommand{\calP}{\mathcal{P}}
\newcommand{\calS}{\mathcal{S}}
\newcommand{\calM}{\mathcal{M}}

\newcommand{\Uself}{\mathbf{U}_{\mathrm{self}}}
\newcommand{\Unear}{\mathbf{U}_{\mathrm{near}}}
\newcommand{\Ufar}{\mathbf{U}_{\mathrm{far}}}

\newcommand{\bigO}{\mathcal{O}}

\newcommand{\figref}[1]{Fig.~\ref{#1}}

\newcommand{\secref}[1]{Section \ref{#1}}

\newcommand{\uback}{\mathbf{u}_{\infty}}

\newcommand{\uhat}{\widehat{\mathbf{u}}}

\usepackage{graphicx}
\usepackage{newtxtext}
\usepackage{newtxmath}
\usepackage{natbib}
\usepackage{hyperref}
\hypersetup{
    colorlinks = true,
    urlcolor   = blue,
    citecolor  = black,
}

\newcommand{\RomanNumeralCaps}[1]
\linenumbers

\title{VesNet: Neural network accelerated solver for simulating Stokesian vesicle suspensions}

\author{S. Zhong\aff{1},
  G. Kabacao\u{g}lu\aff{2}
\corresp{\email{gokberk.kabacaoglu@durham.ac.uk}},
 \and G. Biros\aff{1,3}}

\affiliation{\aff{1} Oden Institute for Computational Science and Engineering, The University of Texas at Austin, Austin, TX-78712, USA
\aff{2}Department of Computer Science, Durham University, Durham, DH1 3LE, UK
\aff{3} Walker Department of Mechanical Engineering, The University of Texas at Austin, Austin, TX-78712, USA }

\begin{document}
\maketitle

\begin{abstract}
Numerically simulating suspensions of deformable particles in the Stokes flow regime is computationally demanding due to strongly nonlinear fluid–structure interactions, evolving interfaces, and multiscale hydrodynamic phenomena. 
We introduce \vesnet, a framework developed to significantly accelerate simulations of vesicle suspensions in two dimensions. Vesicles are deformable particles that maintain local surface inextensibility and resist bending.  \vesnet comprises  components that approximate vesicle self-interactions, including coupling to the imposed background flow and short-range lubrication forces that arise in near-contact situations. Instead of acting as a purely neural surrogate model, \vesnet is a hybrid approach that also retains conventional algorithmic modules for boundary reparameterization and N-body far-field hydrodynamic interactions. A GPU-accelerated implementation of \vesnet delivers more than a 100$\times$ speedup over a multithreaded MATLAB CPU implementation of a state-of-the-art, fully resolved boundary integral equation (BIE) solver, and roughly a 5$\times$ speedup relative to a GPU implementation of the same BIE solver. 
To evaluate its fidelity, we analyze the recovery of single-vesicle phase diagrams, the dynamics of interacting vesicle pairs, and large-scale simulations with thousands of vesicles in Taylor–Green and Poiseuille flow settings. \vesnet successfully reproduces the main observables of interest, thereby enabling extensive simulations using relatively modest computational resources.
\end{abstract}



\section{Introduction}
Particulate suspensions play a vital role in diverse domains, from biological systems \citep{dutta_shelley_e24} and materials science \citep{cademartiri2015programmable}, to advanced manufacturing \citep{haley2019modelling}, and soft robotics \citep{ali2017bacteria}. These systems often operate at low Reynolds numbers, where viscous forces dominate, placing their dynamics within the Stokes flow regime. Modeling such systems is challenging due to the complex, multiscale hydrodynamic interactions, which exhibit long-range many-body effects and short-range lubrication forces \citep{quaife-biros14,rahimian-biros-e15,agarwal-biros20,freund14,maxey2017simulation,stein2024computational}. 

\begin{figure}
    \centerline{
    \includegraphics[width=1\textwidth]{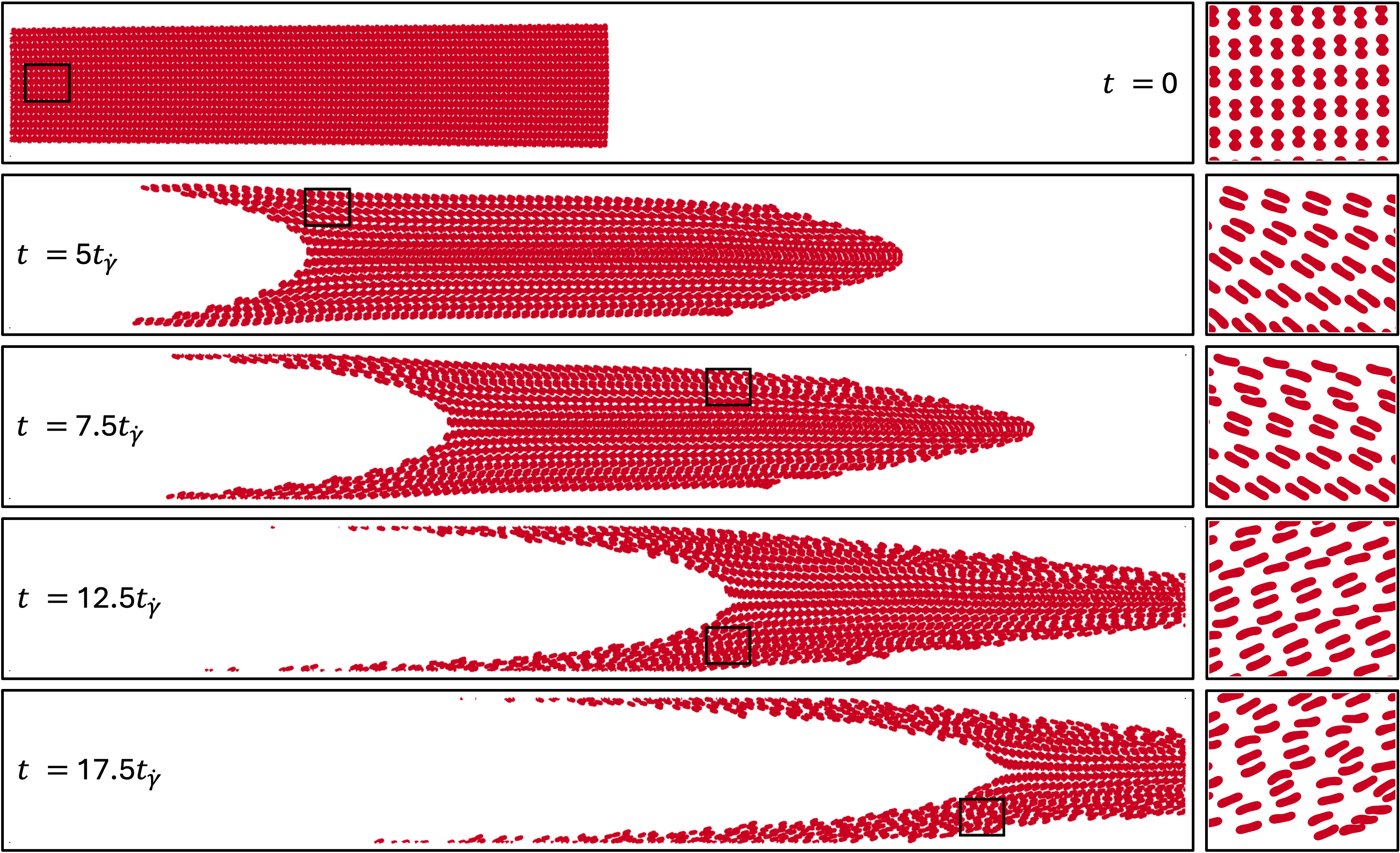}}
    \caption{2,000 vesicles in Poiseuille flow (see sections \ref{sec:pois} and \ref{sec:denseSuspension} for details). The characteristic flow time scale is $t_{\dot{\gamma}} = 1/\dot{\gamma}$, with $\dot{\gamma} = u_{\max}/W$, where $u_{\max}$ denotes the peak velocity at the channel center and $W$ is the channel width. The initial vesicle area fraction is 30\%. Under the imposed flow, vesicles disperse and interact with each other. Each row shows the suspension at successive times. The left column displays the full suspension, while the right column provides magnified views of the regions enclosed by the black rectangular windows in the left column. \vesnet combines several neural networks for self-interactions, advection, and near-singular evaluation, together with N-body interactions and spectral operators for computing elastic forces, performing interpolation, and enforcing area and length conservation. The complete simulation is executed on a GPU for 4000 steps with a time step of $\Delta t = 10^{-5}$. A single \vesnet time step for this simulation takes 1.4 s on one NVIDIA GH200.
    }
    \label{fig:2000Parabolic}
\end{figure}

We demonstrate \vesnet's computational accuracy and performance on a specific particulate flow: vesicle flows (see \figref{fig:2000Parabolic}). Vesicles are highly deformable yet inextensible membranes containing a Newtonian fluid \citep{seifert97}. They resist  stretching, compression, and bending, and have nearly constant surface area and enclosed fluid volume. Vesicles serve as numerical and experimental proxies of red blood cells \citep{fedosov-gompper-e14,misbah12,kabacaoglu-biros19a,reichel-fedosov-e19}. Fast and accurate simulations of vesicle suspensions provide an indispensable tool to discover optimal designs of microfluidic devices in medical applications and beyond \citep{kabacaoglu-biros18b, liu2025data}.   

\subsection{Contributions}
\vesnet stands out distinctly from existing particulate flow simulation methods  based on machine learning (ML) due to its integration of neural networks (NN) with the boundary integral equation method (BIEM) and its focus on deformable particles. Building on MLARM \citep{kabacaoglu-biros19pre}, we enhance both accuracy and performance to enable dense vesicle suspension simulations with highly deformable vesicles. The resulting ML-accelerated solver, \vesnet, features:
\begin{itemize}
    \item \vesnet approximates the integro-differential operators using the Iterated V-shape Net (\ivnet) introduced by \citep{zhong2026ivnetneuralnetworkelliptic}. \ivnet captures the fact that the solution to the Stokesian particulate flows involves convolutions with the fundamental solution to the Stokes equations and has a multiscale nature. This network architecture enables accurate approximations of the nonlinear operators.

    \item \vesnet uses an an \ivnet to handle nearly-singular integrals for vesicles in close contact to enable dense suspension simulations and introduces an algorithm to efficiently couple it to arbitrary suspensions. 

    \item \vesnet is fully GPU accelerated; all steps, including solver steps and the network inference run on a GPU.  This enables fast simulations of vesicle flows, offering a robust tool for scenarios such as parameter exploration, optimization problems, and uncertainty quantification where traditional simulations and existing ML methods fall short. The code is open source and released at \url{https://github.com/bioflume/Ves2Dpy}. 
\end{itemize}

\vesnet embeds physical invariances—such as translation, rotation, and scaling—directly into both its training data and architecture, resulting in a generalizable and interpretable model. This design enables \vesnet to achieve BIEM-level accuracy at reduced spatial and temporal resolutions.

\subsection{Related work}

\textit{High fidelity methods for suspensions of deformable particles.} Simulation methods for particulate flows typically fall into two categories. First, methods based on solvers for  boundary value problems,  such as BIEM \citep{shravan-biros-e09} and immersed boundary methods \citep{peskin2002immersed}, require solving large linear systems at every time step, making them computationally expensive for large particle counts. Second, particle methods, including Stokesian Dynamics \citep{brady1988stokesian}, Smoothed Particle Hydrodynamics \citep{monaghan2012smoothed}, and Dissipative Particle Dynamics \citep{pivkin2010dissipative, bian2012multiscale}, bypass PDE solutions but still involve intensive computations like matrix inversions, especially for large-scale systems.  Of course, this list is not exhaustive, and other methods (e.g., Lattice Boltzmann) and hybrid versions exist. \todo{this needs more work}

In this study, we present \textit{\vesnet}, a ML-accelerated solver that synergistically integrates ML algorithms with BIEM to address the inherent computational bottlenecks of conventional BIEM frameworks. This hybrid approach enables rapid and precise simulations of two-dimensional (2D) dense particulate Stokes flows.
The methodology developed here establishes a foundational framework for future extension to 3D systems, paving the way for high-performance simulations at previously unattainable scales.
%

\textit{Machine learning algorithms for suspensions.} Recent advances in ML have opened transformative avenues for modeling complex physical systems, particularly those involving interacting particles. ML is used for approximating force fields \citep{zhang2018end}, deriving coarse-grained models \citep{majewski2023machine}, approximating hydrodynamic interactions \citep{pfaff2020learning} and enabling multiscale modeling \citep{wang2020machine}. Most particulate flow simulations using ML consider rigid particles. While a vast body of research exists on this topic, we refer to the recent studies to highlight the trends \citep{aminimajd2025robust,aminimajd2025scalability}. One of the most recent studies is NeuralDEM \citet{alkin2024neuraldem}. It replaces traditional, computationally intensive discrete element method (DEM) simulations with fast, adaptable ML surrogates. This model captures long-term transport processes across various regimes using macroscopic observables, eliminating the need for detailed microscopic parameters. Another study for DEM is by \citet{xu2022improved} where ML is integrated into DEM through the Newton's second law. The model addresses the complex behavior of granular materials, providing efficient and reliable predictions for various applications.

Recent works have used graph neural networks (GNNs) to simulate deformable particles where a drop is represented as a group of rigid particles constrained to move collectively \citep{ma2022fast, ma2024shape}. GNNs consider particulate systems as dynamic graphs, while the vertices correspond to particles and graph edges capture pairwise interactions. At each time step vertex latent features are updated by reduction over edges. The deeper the GNN the longer the interactions. 
This architecture enables trained models to generalize to systems with arbitrary particle numbers. The advantage of GNNs is that they capture the dynamics of the entire suspension. Their disadvantage is that require large training sets that depend on the suspension conditions, e.g., the background forcing flow. 

\textit{Boundary integral operator approximation.} By this we refer to deep learning approximation of linear and nonlinear boundary integral operators. One class of methods for such purpose are neural operators~\citep{liFourierNeuralOperator2021, kopanicakovaDeepOnetBasedPreconditioning2024, oleary-roseberryDerivativeInformedNeuralOperator2024, luLearningNonlinearOperators2021, caoChooseTransformerFourier2021, bhattacharya_model_2021}. In our context, we apply neural operators to approximate  flow maps and boundary integral operators. Although there are many alternatives, we use \ivnet as the basic architecture. In fact \ivnet  is similar to the original U-Net architecture \citep{ronneberger2015u}, a widely-used convolutional neural network known for its encoder-decoder structure and multiscale feature extraction capabilities. 

Besides \ivnet, the overall structure of \vesnet is based on our previous work \citep{kabacaoglu-biros19pre}. In that paper,
we introduced the basic idea of training on  a single vesicle configuration evolved on short-time horizons. The network architecture in \citep{kabacaoglu-biros19pre} was quite simple, just fully connected perceptron layers. Furthermore, it was restricted to dilute suspensions due to the lack of approximations for near singular interactions.


\subsection{Limitations} 
\vesnet currently does not accommodate viscosity contrasts between the interior and exterior fluids of a vesicle. In the case of viscosity contrast the boundary integral formulation involves another integro-differential operator which can also be approximated in the same manner done in \vesnet. Another limitation is that our work is in 2D.
We remark that vesicle flows in 2D capture many interesting physical phenomena \citep{fedosov-gompper-e12,fedosov-gompper-e14,kaoui-misbah-e11,freund14}.
While extending our framework to 3D poses several challenges, conducting this study in 2D provides valuable intuition on possible issues and ways to overcome them such as data standardization, network architecture, and designing the hybrid solver. \vesnet currently cannot guarantee collision-free vesicle-vesicle interactions which are difficult even in the BIEM. While an approach exists that requires solving a linear complementarity problem with the integral equations for the vesicle flows \citep{lu-zorin-e17}, incorporating this approach into \vesnet would be computationally expensive. We leave the extension of \vesnet to 3D and designing an ML-based collision handling scheme for future work.


\subsection{Outline} 
The paper is organized as follows. In \secref{sec:formulation} we present the governing equations for vesicle flows and our in-house BIEM to solve those equations. In \secref{sec:mlsolver} we explain the proposed solver, \vesnet. Its overall algorithm and GPU implementation are in \secref{sec:vesnet}. Finally, we show the accuracy and computational performance of \vesnet in \secref{sec:results}.

\begin{table}
\begin{center}
\begin{tabular}{cccc} \hline
 \bf Symbol & \bf Definition & \bf Function & \bf Definition\\
         \hline
         $\Omega$ & Fluid domain & $\GG(\rr)$ & Stokeslet, \eqref{eq:Stokeslet}\\
         $\xx$ & Point in $\Omega$  & $\calS[\XX,\ff](\xx)$ & Single layer integral (SLI) \eqref{eq:single_layer}   \\
         $\uu$ & Velocity & $\calB(\XX)(\xx)$ & Velocity due to bending force \eqref{eq:bending_op}  \\
         $\mu$ & Viscosity & $\calT(\XX,\xx)\sigma$ &  Velocity due to tension \eqref{eq:tension_op}\\
         $\gamma$ & Vesicle membrane  & $\calP(\XX)$  & Surface divergence \eqref{eq:inextensib} \\
         $\XX$ & Point on $\gamma$  & $\calL(\XX)$ & Inextensibility \eqref{eq:PTop} \\
         $s$ & Arclength parameter  & $\calM(\XX)$ & $ \calT(\XX,\XX)\calL^{-1}(\XX)\calP(\XX)$ \\
         $\sigma$ & Membrane tension  & $S_{\uu}^k$ & Map of $\XX$ to velocity-related tension term \eqref{eq:advTenMap} \\
         $\ff$ & Membrane elastic force  & $X_{\uu}^k$  & Map of $\XX$ to advection term \eqref{eq:advectionMap}\\
         $T$ & Simulation time horizon  & $S_{\sigma}$  & Map of $\XX$ to self-tension term\\
         $N$ & Number of points on $\gamma$  & $X_{\XX}$  & Relaxation flow map  \\
         $M$ & Number of vesicles  & $N_{\XX}^k$ & Map of $\XX$ to SLI of Fourier basis in near-field \eqref{eq:NearFieldMap} \\
         \hline
\end{tabular}
\caption{Index of frequently used symbols and operators.}\label{tab:symbolsIndex}
\end{center}
\end{table}




\section{Problem formulation}\label{sec:formulation}
In this section, we present the governing equations for vesicle flows in the limit of vanishing Reynolds numbers and introduce their boundary integral equation formulation. The list of symbols and their definitions are in Table \ref{tab:symbolsIndex}.

\begin{figure}
    \centering
    \includegraphics[width=0.95\textwidth]{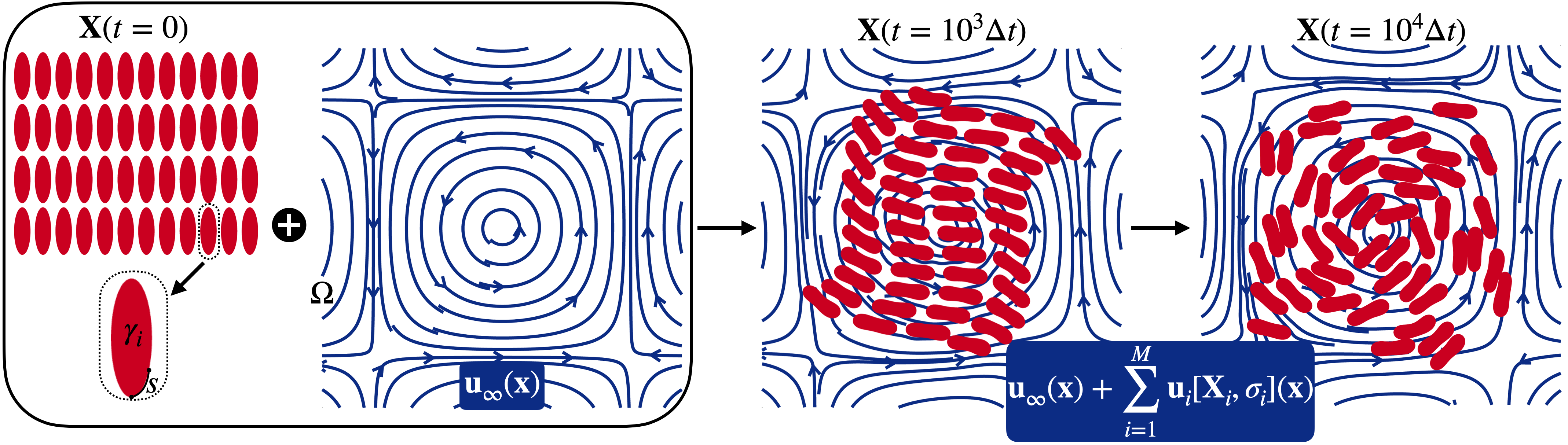}
    \caption{Vesicle suspension in a free-space Taylor-Green flow. The left figure shows the initial configuration of 48 vesicles (at $t = 0$) in red. Each vesicle membrane is denoted by $\gamma_i$, and its discretized points are represented by $\XX$. The membrane is parameterized by the arclength parameter $s$. The domain is denoted by $\Omega$, with $\xx$ representing a point within it. The vesicles move under the influence of the background flow $\uback(\xx)$, which, in this case, follows a Taylor-Green vortex, with streamlines shown in blue. As they move, the vesicles interact with one another through the suspending fluid. The middle figure displays the vesicle configurations after $10^3 \Delta t$ time steps, superimposed with the streamlines of the total flow. The total flow consists of both the background flow $\uback(\xx)$ and the flow induced by the vesicles due to their elastic forces on the fluid, represented as $\uu_i [\XX_i, \sigma_i](\xx)$. The membrane’s elastic force has two components: a bending force, given by $\kappa_b \XX_{ssss}$ where $\kappa_b$ is the bending stiffness, and a tensile force due to membrane tension $\sigma$. The right figure shows the vesicle configurations after $10^4 \Delta t$time steps, illustrating their long-term evolution under the flow.}
    \label{FigSchematic}
\end{figure}

\subsection{Vesicle flows}

Consider multiple vesicles (see \figref{FigSchematic}) suspended in a Newtonian fluid moving with a background flow, $\uback (\xx)$. Let $\gamma_i$ denote the membrane of a vesicle discretized in arclength $s$. Then, $\{\gamma_i\}_{i=1}^M$ is a collection of vesicle membranes and $\gamma = \bigcup_{i=1}^{M}\gamma_{i}$. The points $\XX_i(\alpha,t)$ on $\gamma_i$ are given at $N$ uniformly distributed points $\{ \alpha_k = 2\pi k/N\}_{k=1}^N$ in the parametric domain $\alpha$ and time $t$. In the limit of vanishingly small Reynolds number, the vesicle flow is governed by the coupled fluid-solid interaction equations
\begin{align}
  -\nabla p(\xx) + \mu \Delta \uu(\xx) &= 0,  \quad & \xx \in \Omega \setminus \gamma,  \quad & \text{conservation of momentum, }&\label{eq:Stokes} \\
  \nabla \cdot \uu (\xx) &= 0, \quad & 
 \xx \in \Omega \setminus \gamma, \quad & \text{conservation of mass,}& \label{eq:incomp} \\
  \uu(\xx) & = \uback(\xx), \quad & \xx \rightarrow \infty, \quad & \text{free-space boundary condition,}& \label{eq:freeSpace} \\
  \XX_s \cdot \uu_s & = 0, \quad & \XX \in \gamma, \quad & \text{vesicle inextensibility,}& \label{eq:inextensib} \\
  \uu(\XX,t) &= \dot{\XX}, \quad & \XX \in \gamma,  \quad & \text{velocity continuity,}& \label{eq:vesCont} \\
  \ff(\XX) & = [\![ T \nn(\XX) ]\!], \quad & \XX \in \gamma, \quad & \text{traction jump.}& \label{eq:tracJump}
\end{align}
Here, $\uu$ is velocity, $p$ is pressure, $\mu$ is viscosity, $T = -p + \mu (\nabla \uu + \nabla \uu^T)$ is the Cauchy stress tensor, and $\nn$ is the outward normal vector the vesicle membrane at point $\XX$. $[\![ \cdot ]\!]$ denotes the jump accross the membrane. $\dot{\XX}$ is the velocity of the points on the vesicle membrane. 

We assume that the fluid inside the vesicles is identical to the suspending fluid. The vesicle membrane is inextensible and resists stretching due to hydrodynamic forces \citep{seifert97, kantsler-steinberg-e05, misbah12}. Hence, the membrane applies force $\ff$ to the surrounding fluid that balances the hydrodynamic stress jump across the vesicle membrane
\begin{equation}\label{eq:elastic_force}
    \ff (s,t) = -\kappa_b \XX_{ssss} + (\sigma \XX_s)_s
\end{equation}
where $\kappa_b$ is the bending stiffness, $(\cdot)_s$ is the derivative with respect to arclength, and $\sigma$ is the membrane tension. This force is given by the functional derivative of the Helfrich bending energy \citep{Zhong-Helfrich89,kaoui-zimmermann-08} $E = \frac{\kappa_b}{2}\oint c^2 \, ds + \oint \sigma \, ds$ where $c$ is the membrane curvature and tension $\sigma$ is a Lagrange multiplier enforcing the local arclength conservation constraint (inextensibility).


\subsection{Boundary integral equation method}
Given membrane configuration $\XX$ and tension $\sigma$ on vesicle $\gamma$, the integral equation formulation gives the velocity at $\xx$ due to that vesicle as the single layer integral $\calS$ \citep{rallison-acrivos78,pozrikidis92,pozrikidis01a}
\begin{equation}\label{eq:single_layer}
    \calS[\XX,\ff](\xx) =\int_{\gamma}\GG(\xx-\XX) \ff(\XX) \, ds(\XX),
\end{equation}
where $\GG$ is a fundamental solution to the Stokes equations (stokeslet) in two dimensions 
\begin{equation}\label{eq:Stokeslet}
 \GG(\rr) =  \frac{1}{4\pi \mu} \left(\log |\rr| + \frac{\rr \otimes \rr}{| \rr |^2}\right), \quad \rr = \xx - \XX.
\end{equation} 

Solving \eqref{eq:Stokes}-\eqref{eq:tracJump} with the Boundary Integral Equation Method amounts to solving two integro-differential equations for every vesicle for the unknowns $\XX$ and $\sigma$.
\begin{align}
  &\begin{aligned}\label{eq:membrane_velocity}
       \frac{\partial \XX_i}{\partial t} = \Uself[\XX_i, \sigma_i] + \Unear[\overline{\XX}, \overline{\sigma}](\XX_i) + \Ufar [\overline{\XX}, \overline{\sigma}](\XX_i) + \uback(\XX_i),
  \end{aligned}\\
  &\begin{aligned}\label{eq:membrane_tension}
    \calL(\XX_i)\sigma_i = -\calP(\XX_i)\biggl(\calB(\XX_i)\XX_i + \Unear[\overline{\XX}, \overline{\sigma}](\XX_i) + \Ufar [\overline{\XX}, \overline{\sigma}](\XX_i) + \uback(\XX_i)\biggr),
  \end{aligned}
\end{align}
for $i = 1, \cdots, M$. Equation \eqref{eq:membrane_velocity} is to evolve the membrane positions $\XX_i$ where $(\overline{\XX}, \overline{\sigma})$ denote the membranes and the tension of the other vesicles in the suspension, $\Uself$, $\Unear$, and $\Ufar$ refer to the velocity induced by the vesicle itself, the vesicles in the near-field of the $i^{\mathrm{th}}$ vesicle and those in its far-field, respectively. A vesicle is considered in the near-field of the point $\xx$ if the distance of $\xx$ to the vesicle, ${\text{inf}}_{\XX \in \gamma} \| \xx - \XX\|_2$, is less than or equal to the arclength spacing on the vesicle membrane. Equation \eqref{eq:membrane_tension} is for the tension $\sigma_i$.

Here, the symbols are defined as 
\begin{subequations}
\begin{align}
    & \boldsymbol{U}[\XX,\sigma](\xx) = \calB(\XX)\xx + \calT(\XX,\xx)\sigma \label{eq:totVelocity} \\
    & \calB(\XX)\xx = -\kappa_b \calS[\XX,\XX_{ssss}](\xx), \label{eq:bending_op} \\
    & \calT(\XX,\xx)\sigma = \calS[\XX,(\sigma\XX_s)_s](\xx), \label{eq:tension_op} \\
    & \calL(\XX) = \calP(\XX)\calT(\XX,\XX). \label{eq:PTop}
\end{align}
\end{subequations}
The velocity $\boldsymbol{U}$ represents the velocity due to bending $\calB(\XX)\xx$ and the velocity due to tension $\calT(\XX,\xx)\sigma$. For the remainder of the manuscript, we adopt a shorthand notation for ease of reference. We suppress the subscripts $(i, \, j)$ and the unknowns $\XX$ and $\sigma$ in parentheses:
\begin{align*}
    &\Uself = \Uself [\XX_i, \sigma_i], \quad \Unear = \Unear[\overline{\XX}, \overline{\sigma}](\XX_i), \quad \Ufar[\overline{\XX}, \overline{\sigma}](\XX_i), \\
    &\calB \XX = \calB(\XX)\XX, \quad \calT \sigma = \calT(\XX,\XX)\sigma, \quad \calL = \calL(\XX), \quad \calP = \calP(\XX), \quad \uback = \uback(\XX).
\end{align*}

\subsubsection{Spatial discretization}
We use a Lagrangian formulation where we track the position of material points on $\gamma$ discretized using spectral discretization. Let $\{\alpha_k = 2k\pi/N \}_{k=1}^N$ be $N$ material points, then the membrane points are 
\begin{equation*}
    \XX(\alpha) = \sum_{k = -N/2+1}^{N/2} e^{ik\alpha} \widehat{\XX}_k
\end{equation*}
where $\widehat{\XX}_k$ is the Fourier coefficient corresponding to the $k^{\text{th}}$ mode. In the ground truth simulations shown in \secref{sec:results}, we use large number of discretization points ($N = 128$) and maintain equal spacing of the points along vesicle membranes. We compute arclength derivatives with spectral accuracy accelerated with Fast Fourier Transform 
\begin{equation*}
    \frac{\partial }{\partial s} = \frac{\partial \alpha}{\partial s}\frac{\partial }{\partial \alpha} = \frac{1}{\| \partial \XX / \partial \alpha\|}\frac{\partial }{\partial \alpha}.
\end{equation*}

\subsubsection{Singular and near-singular integration}\label{sec:nearsing-hedge}
We use Nystr\"{o}m-type quadrature rules to discretize integral operators. Different quadrature rules are used to maintain uniform accuracy in calculating the single layer integral for the near-field and the far-field. For target points in the far-field, the trapezoid rule approximates the integral with high accuracy. Evaluating the integral for the same source and target points suffers from a logarithmic singularity which we remove by using the hybrid Gauss-trapezoid quadrature rule \citep{alpert99}. The direct evaluation of the single layer integral using this quadrature rule has $\bigO(N^2 + Nm)$ complexity which is reduced to $\bigO(N+Nm)$ with fast multipole method (FMM) \citep{greengard-rokhlin87,ying-zorin-e03}. We refer the reader to \citet{quaife-biros14} for further details.

For the near-field, we use the efficient and accurate method presented in \citet{ying-zorin-e06,quaife-biros14} (see the middle figure in \figref{fig:near-singular}). Simply put, the idea is (i) to find the projection of the target point on the membrane, then evaluate the integral on the membrane with the singular quadrature rule; and (ii) to place interpolation points in the far-field along the normal to the membrane, then evaluate the integral using the trapezoid rule. Finally, the integral evaluation at the target point is found by interpolating those integral values along the line. If there are $M$ vesicles and $N$ points per vesicle, this algorithm requires $\bigO(MN)$ operations for the trapezoid rule accelerated with FMM. Besides, this method requires the singular quadrature for the interpolation point on the membrane.  While the singular integration in 2D is not a bottleneck, a state-of-the-art singular integration scheme for spherical harmonics accounts for approximately 20\% of the total time in 3D \citep{malhotra-biros-e17}.

\subsubsection{Discretization in time}
\label{sec:biem-time-discretization}
The leading source of numerical stiffness in vesicle simulations is $\calB_i$ which results in a time step restriction $\Delta t \sim h^3$ \citep{quaife-biros15}. As explicit time stepping schemes suffer from the stiffness, variants of semi-implicit time stepping schemes are used to remove the time step restrictions \citep{shravan-biros-e09,rahimian-biros-e10,quaife-biros14}. In our ground truth simulations, we use a fully implicit time stepping scheme \citep{quaife-biros14}. Let $\XX^n$ and $\sigma^n$ be the membrane configuration and tension at the current time step. The integro-differential operators are discretized at $\XX^n$ while vesicle-vesicle interactions, bending and tension forces are treated implicitly. 
\begin{align}\label{eq:time_step_sc1}
    &\frac{\XX^{n+1} - \XX^n}{\Delta t}  = \Uself^{n+1} + \Unear^{n+1} + \Ufar^{n+1}, \nonumber \\
    &\calL^n\sigma^{n+1} = -\calP^n\left(\calB \XX^{n+1} + \Unear^{n+1} + \Ufar^{n+1} + \uback^n \right).
\end{align}
\eqref{eq:time_step_sc1} is fully coupled and leads to a dense linear system. We solve it with preconditioned GMRES \citep{saad03}. The number of GMRES iterations for this preconditioned linear system depends on the inter-vesicle proximity \citep{quaife-biros14} and is mesh-independent.

\section{Machine-learning-accelerated solver: \vesnet}\label{sec:mlsolver}
In  this section, we present the ML-accelerated solver for vesicle simulations, \textit{\vesnet}. The solver uses NNs to approximate the integro-differential operators \citep{kabacaoglu-biros19pre}. We describe the time-stepping scheme used in the solver in \secref{sec:timeStep} and detail the integral operators in \secref{sec:nnoperators}. We improve the stability and accuracy of the solver for dense suspensions by developing a fast near-singular integration strategy that uses NN approximations instead of a numerical near-singular scheme in \cite{quaife-biros14}. We present the new near-singular integration in \secref{sec:nearsing}. All the networks in the solver take a vesicle configuration as an input and are trained using single-vesicle short horizon simulations. They do not require retraining for different flow conditions (e.g., the number of vesicles and imposed boundary condition). In \secref{sec:architecture} we present the architectures of the neural networks and the training details. An overview of the main components of \vesnet\ is shown in \figref{fig:allNets}.


\subsection{Discretization in time}\label{sec:timeStep}
We want \vesnet to enable modular training and to require only a single vesicle as an input. For that purpose, we treat vesicle-vesicle interactions explicitly and use an operator splitting method \citep{macnamara-strang16} to solve \eqref{eq:membrane_velocity} and \eqref{eq:membrane_tension}. The bending force is still implicitly treated to avoid stringent time stepping constraints. In \vesnet's time stepping, we eliminate the tension $\sigma$ in \eqref{eq:membrane_velocity} by substituting the solution to \eqref{eq:membrane_tension} 
\begin{equation*}
    \sigma = -\calL^{-1}\calP \left( \calB \XX + \Unear + \Ufar + \uback \right)
\end{equation*}
into \eqref{eq:membrane_velocity}
\begin{equation}\label{eq:membraneVelwoTen}
    \frac{\partial \XX}{\partial t} = (1-\calM) \calB \XX + (1-\calM) \left(\Unear + \Ufar + \uback \right),
\end{equation}
where $\calM = \calT \calL^{-1}\calP$ \citep{rahimian-biros-e10}.

Let $\XX^n$ and $\sigma^n$ denote the vesicle configurations and tension in the previous time step. Applying an operator splitting to \eqref{eq:membraneVelwoTen} leads to two subproblems: advection and relaxation. Given $\XX^n$ and $\sigma^n$, we first evolve the vesicles with the velocity induced by the vesicle-vesicle interactions and the background flow by solving the advection problem to find $\XX^*$, i.e.,
\begin{equation}\label{eq:advection_disc}
    \frac{\XX^* - \XX^n}{\Delta t} = (1-\calM^n) \left(\Unear + \Ufar + \uback \right),
\end{equation}
where $\Delta t$ is the time step size. Then we solve the relaxation problem to find the vesicle configurations at the next time step $\XX^{n+1}$
\begin{equation}\label{eq:relaxation_disc}
    \frac{\XX^{n+1} - \XX^*}{\Delta t} = (1-\calM^*) \calB^* \XX^{n+1}.
\end{equation}
Lastly, the tension is updated as
\begin{equation} \label{eq:tension_disc}
    \sigma^{n+1} = -(\calL^n)^{-1}\calP^n \left(\calB\XX^{n+1}\right) - (\calL^n)^{-1}\calP^n\left( \Unear^n + \Ufar^n + \uback^n \right).
\end{equation}


\subsection{Neural network approximations to integro-differential operators} \label{sec:nnoperators}
In \vesnet, we identify computationally expensive integro-differential operators in \eqref{eq:advection_disc}, \eqref{eq:relaxation_disc}, and \eqref{eq:tension_disc} that require fine spatial resolution and depend solely on a single vesicle's configuration. Then, we use NN to approximate their actions.

\eqref{eq:relaxation_disc} depends only on a single vesicle. Hence, we approximate the flow map, $X_{\XX}: \XX^* \rightarrow \XX^{n+1}-\XX^*$. The flow map $X_{\XX}$ is implicit in time step size $\Delta t$ and the bending stiffness $\kappa_b$. Therefore, retraining the network is needed if these parameters change. We train several networks for different values of these parameters and build an interpolation for any values. Additionally, we also approximate the map $S_{\sigma}: \XX^n \rightarrow (\calL^n)^{-1}\left(\calP^n \calB^n \right)\XX^n$ in \eqref{eq:tension_disc}.  

The other term in \eqref{eq:tension_disc} and the advection problem \eqref{eq:advection_disc} involve the vesicle-vesicle interactions and the imposed background flow, i.e., $\ubar = (\Unear + \Ufar + \uback)$. We remove the dependence on the velocity $\ubar$ so that \vesnet generalizes to unseen problem setups. We use the fact that the dependence on the velocity is linear and decompose velocity using an $N_f$-term truncated Fourier series
\begin{equation*}
    \ubar(\XX(\alpha)) = \sum_{k=-N_f/2+1}^{N_f/2} e^{ik\alpha} \uhat_k
\end{equation*}
where $\uhat_k$ is the Fourier coefficent of the $k^{\text{th}}$ mode. Then, let $V$ be any linear operator, we can write
\begin{equation*}
    V[\XX]\ubar = \sum_{k=-N_f/2+1}^{N_f/2} \left(V[\XX] e^{ik\alpha}\right) \uhat_k. 
\end{equation*}
We replace the velocity term with its Fourier expansion in  \eqref{eq:advection_disc} and \eqref{eq:tension_disc}. This allows us to train one network for each of the $N_f$ maps 
\begin{align}
    X^k_{\uu}: & \XX^n \rightarrow \left(1-\calM^n\right) e^{ik\alpha}, \label{eq:advectionMap}\\
    S^k_{\uu}: & \XX^n \rightarrow (\calL^n)^{-1}\calP^n e^{ik\alpha}, \label{eq:advTenMap} 
\end{align}
Given a vesicle $\XX^n$ and velocity $\ubar$, we perform FFT to obtain the Fourier coefficients of $\ubar$ and approximate the maps $X^k_{\uu}$ and $S^k_{\uu}$ with neural networks. Finally, we apply the inverse Fourier transform. These maps depend linearly on parameters such as the membrane rigidity and the time-step size, so there is no need to retrain the networks for these maps.

\begin{figure}
    \centering
    \includegraphics[width=0.95\textwidth]{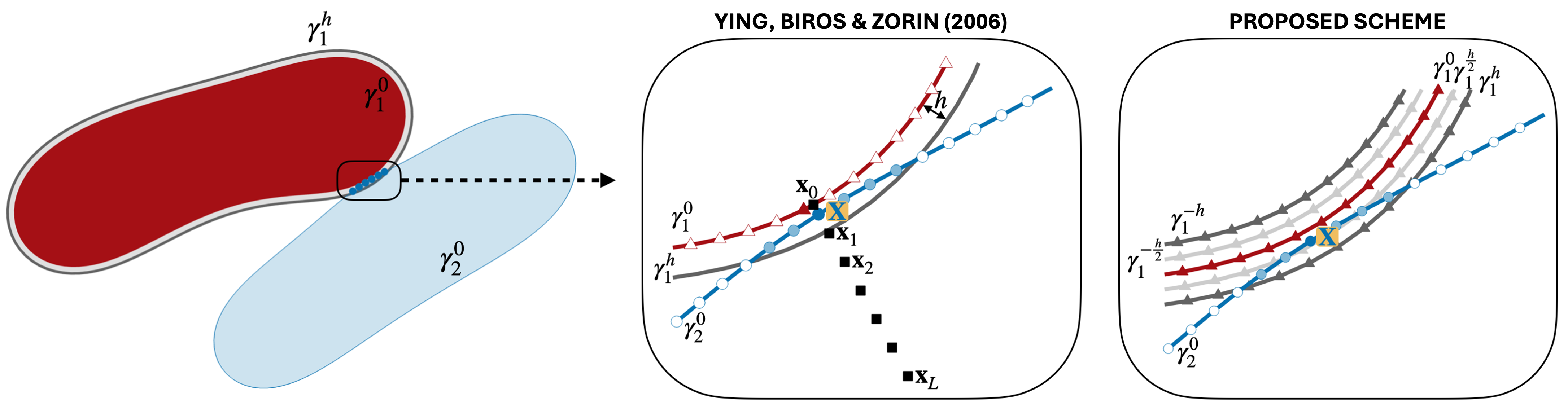}
    \caption{Near-singular integration schemes with the state-of-the-art numerical method (in the middle) and the proposed neural network approach (on the right). On the left figure, two vesicles are in a free-space flow. $\gamma_1^0$ and $\gamma_2^0$ denote the membranes of the red and the blue vesicles, respectively. The gray line denoted with $\gamma_1^h$ defines the boundary of the red vesicle's near-field. The blue circles are the discretization points on $\gamma_2^0$ that are inside the red vesicle's near-field. Hence, the single layer integral due to $\gamma_1^0$ at these points is nearly singular. Let us choose a specific target point $\XX$ (the darker blue point in the middle and the right figures) to explain both near-singular integration schemes. The triangular and circular markers represent the discretization points on $\gamma_1^0$ and $\gamma_2^0$, respectively. In \citet{ying-zorin-e06} the integral is accurately evaluated at the point $\xx_0$ on $\gamma_1^0$ that is the closest to $\XX$ using a singular quadrature and also at points $\{\xx_i\}_{i=1}^L$ outside the near-field through a non-singular quadrature by upsampling the number of source points to $N^{3/2}$. Here, the red-filled triangle is the discretization point that is the closest to $\XX$. The single layer integral at $\xx_0$ is obtained through interpolation along the membrane when $\xx_0$ does not align with a discretization point. Then, an interpolation of the the integral values at $\{\xx_i\}_{i=0}^L$ delivers the integral at $\XX$. Our proposed scheme (on the right) removes the necessity to the direct evaluation of the single layer integral outside near-field. Instead, it builds five layers, $\gamma_1^{-h}$ at $d = -h$, $\gamma_1^{-h/2}$ at $d = -h/2$, $\gamma_1^0$ (on the vesicle), $\gamma_1^{h/2}$ at $d = h/2$ and $\gamma_1^{h}$ at $d = h$. At the discretization points (the triangular markers) on these layers the single layer integral is approximated through neural networks. Finally, the integral at $\XX$ is found by interpolating the integral values on those layers.}
    \label{fig:near-singular}
\end{figure}


\subsection{Near-singular integration with neural networks}\label{sec:nearsing}
For accurate and stable simulations with \vesnet, we propose a novel near-singular integration scheme using NNs (see \figref{fig:near-singular}). One of the main advantages of the new scheme is that it does not require singular quadrature for the single layer integral during a simulation. Let $d = {\text{inf}}_{\XX \in \gamma} \| \xx - \XX\|_2$ be the distance of the target point $\xx$ to a vesicle $\gamma$. We consider $|d| \leq h$ as the near field of the vesicle where $h$ is the arc-length spacing. $|d| > h$ is the far field. For the near field, we use NNs to approximate the single layer integral around a vesicle. We consider target points on five layers (see \secref{sec:appendixNear} for a further explanation on the choice of five layers). One of the layers is the vesicle membrane on which the points are distributed equally in arclength . The other four layers are obtained by shifting these points in the normal direction by $-h$, $-h/2$, $h/2$ and $h$. Let $\xx$ denote these target points. Since the single layer integral depends on the elastic force $\ff$ which varies during a simulation, we remove the dependency on $\ff$ by decomposing $\ff$ using Fourier expansion:
\begin{equation*}
    \ff = \sum_{k = -N_f/2}^{N_f/2}  \hat{\mathbf{f}}_k e^{ik\alpha},
\end{equation*}
where $\hat{\mathbf{f}}_k$ is the $k^{th}$ Fourier coefficient. Then, the single layer integral becomes
\begin{equation*}
    \calS[\XX, \ff](\xx) = \sum_{k=-N_f/2}^{N_f/2}  \left( \int_{\gamma} \GG(\xx- \XX) e^{ik\alpha} d\gamma(\XX) \right) \hat{\mathbf{f}}_k.
\end{equation*}
We train neural networks to approximate the maps between a vesicle configuration $\XX$ and its single layer integral with the Fourier basis functions $\calS[\XX, e^{ik\alpha}]\xx$ evaluated at $\xx$ using the near-singular integration scheme \citep{ying-zorin-e06}
\begin{equation}\label{eq:NearFieldMap}
    N_{\XX}^k: \XX \rightarrow \calS[\XX, e^{ik\alpha}]\xx.
\end{equation}
During a simulation, we construct $\calS[\XX, \ff]\xx$ on the layers using the Fourier coefficients $\hat{\mathbf{f}}_k$. With the single layer integral evaluations on the layers, we interpolate the single layer integral at arbitrary points in the near-field. 

We use radial basis function (RBF) interpolation to approximate velocities at arbitrary points in the near zone given the velocities along the five layers predicted by \vesnet. The Gaussian function is selected as the RBF to construct symmetric positive definite kernels, leading to a linear systems for each vesicle. Each of these kernels is of size $5N \times 5N$ with $N$ being the number of discretization points on a vesicle, 5 being the number of layers. The linear systems are efficiently solved by batched Cholesky decomposition in PyTorch. Due to batching, solving RBF coefficients scales sub-linearly with the number of vesicles, with a per-vesicle complexity of $\bigO(N^3)$. Furthermore, the interpolation errors using Gaussian kernels exhibit exponential convergence with respect to resolution refinement. Finally, inferring the velocities at arbitrary points with the computed RBF coefficients requires computing the kernel between query coordinates and the five layers, and performing matrix multiplication between the kernel and RBF coefficients. This leads to a complexity linear to the number of query points. 

\subsection{Data and network architecture}\label{sec:architecture}

In this subsection, we explain the preparation of the input to the networks, namely the standardization step considering the translation, rotation, and scaling invariance of the approximated maps. Then we detail the architecture of the networks used in \vesnet.

\subsubsection{Data representation}
The input to the networks is a vesicle configuration $\XX$. We standardize an input vesicle so that it centers at $(x, y) = (0, 0)$, and its principal axis aligns with the $y$-axis. The center of a vesicle $\XX$ is calculated as 
\begin{equation*}
    \mathbf{c} = \frac{\int_{\gamma}\XX(\XX\cdot \nn)\,d\gamma}{\int_{\gamma}\XX\cdot \nn \, d\gamma}.
\end{equation*}
The principal axis of the vesicle is the axis corresponding to the smallest principal moment of inertia. The moment of inertial tensor $J$ is 
\begin{equation*}
    J = \int_{\omega} \left(\|\rr\|^2 I - \rr \otimes \rr \right) \, d\XX = \frac{1}{4} \int_{\gamma} \rr \cdot \nn \left(\|\rr\|^2 I - \rr \otimes \rr \right) \, ds
\end{equation*}
where $\rr = \XX - \mathbf{c}$, $\XX$ is the point on a vesicle and $\mathbf{c}$ is the center of the vesicle. The trained networks take in a standardized configuration, computes the output, and the output is then de-standardized to be used in the simulation. We also standardize an input vesicle to have discretization points that are distributed equally in arclength. The $N$ discretization points are ordered such that the first point lies on or just above the x-axis, with the remaining points indexed counterclockwise. 

Besides, the maps $X_{\XX}$ and $S_{\sigma}$ scale with the membrane arclength linearly and inversely quadratically, respectively. Hence, we scale vesicles to have a unit length and adjust the network approximations  based on their scaling with the arclength. The maps also depend on the reduced area of a vesicle (i.e., the ratio of the area enclosed by a vesicle to the area of a disk having the same perimeter as the vesicle). Here, we opt for the reduced area of 0.65 which corresponds to healthy red blood cells' reduced volume \citep{mohandas-evans94, tomaiuolo14}. For \vesnet to be used for a different reduced area, its networks must be trained for that reduced area value.


We create the training dataset by performing simulations of free-space/confined fundamental vesicle flows. These simulations include (i) single vesicle in free shear flow at various shear rate values (see \secref{sec:shear}), (ii) single vesicle in free Poiseuille flow (see \secref{sec:pois}) at various shear rate and flowline curvature values, and (iii) dilute vesicle suspensions in a rotating cylinder that contains randomly distributed rigid obstacles of different shapes. These simulations generate a wide range of vesicle configurations, from small to large deformations. To construct a representative dataset with minimal redundancy, we choose distinct vesicles by comparing their geometrical similarity using the Hausdorff distance. Our dataset includes approximately 155,000 vesicle configurations. 
The resolution of the dataset is $N = 32$, where the exact integro-differential maps are built with $N = 128$ points and then downsampled to $N = 32$.

\subsubsection{Network architecture}
The Iterated V-shaped Net (\ivnet) was originally proposed for solving linear, elliptic partial differential equations with varying coefficients. IV-Net imitates an iterative solving process, where each iteration is computed by a multigrid-like network block. The block has a V-shaped structure, performing downsampling and upsampling via direct convolutions in the physical space, so it is essentially a multilevel convolutional network. We choose \ivnet because of its demonstrated power in learning challenging nonlinear solution operators for partial differential equations with high-contrast coefficients, and good performance with few training data. The maps mentioned in \secref{sec:nnoperators} and  \secref{sec:nearsing} are also complicated nonlinear operators that relate vesicle configurations to specific configuration-dependent outputs, making this problem essentially an  operator learning task.

Even though \ivnet was originally proposed for 2D problems, adjusting the architecture for vesicle flow simulations is immediate since we can replace 2D convolutions with 1D convolutions. The most important hyperparameters are: number of blocks $n_b$, channel expansion factor $w$, and downsampling levels $L$. $n_b$ controls how many blocks to use, which can be regarded as the depth of \ivnet. The factor $w$ scales the number of convolutional channels, effectively controlling the network's width. Meanwhile, $L$ determines the downsampling levels by strided convolution, and enables hierarchical processing of information. We perform a grid search for these parameters in order to find the best combination for learning each mapping. 

\begin{figure}
    \centering
    \includegraphics[width=0.98\textwidth]{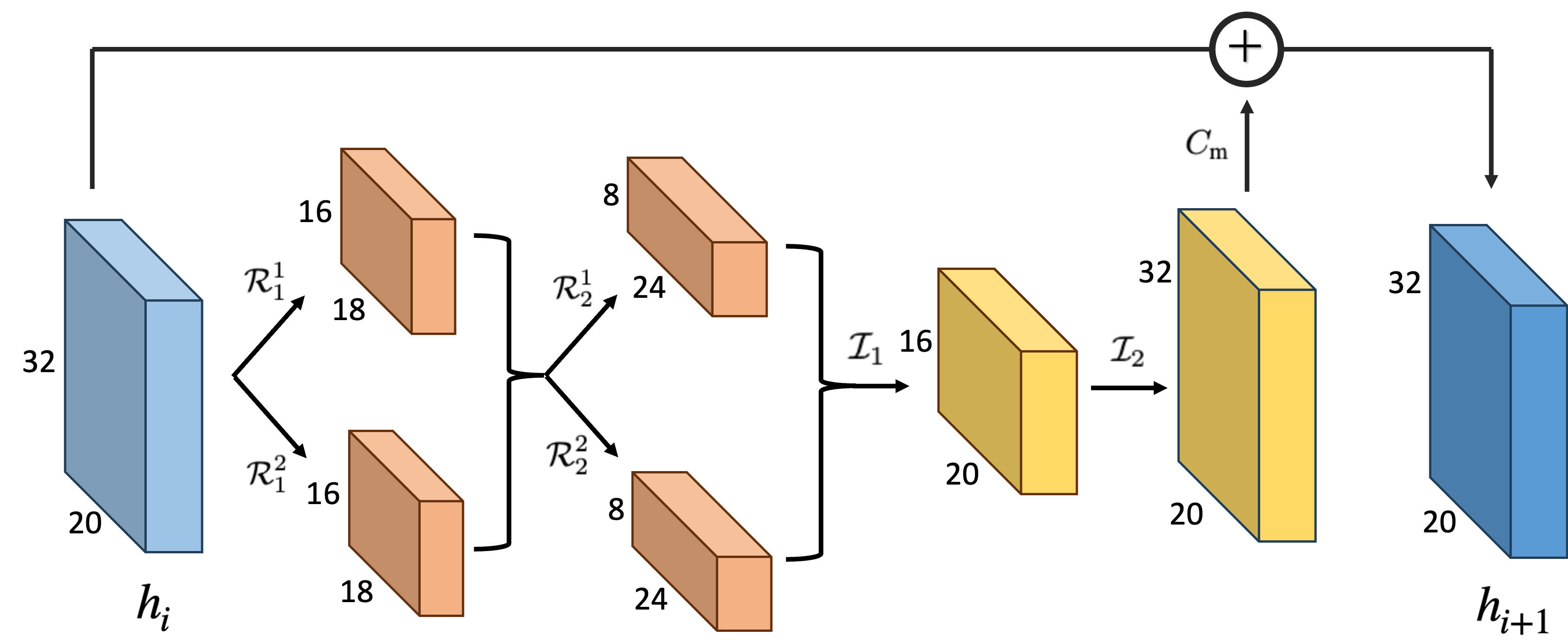}
    \caption{The architecture of a single block in \ivnet for updating 1D hidden representation from $\mathbf{h}_i$ to $\mathbf{h}_{i+1}$. $\mathbf{h}_i$ has $20$ channels and each channel is a vector of length $32$. In the first restriction step, two paths of convolutions with dilation=$1$ and dilation=$2$, denoted by $\mathcal{R}_1^1$ and $\mathcal{R}_1^2$, are used for downsampling to $16$. The outputs from two paths are concatenated in the channel dimension. Similarly, in the second restriction step, apply $\mathcal{R}_2^1$ and $\mathcal{R}_2^2$ to arrive at $8$ (the coarsest level).  Then, apply the interpolation operation twice (denoted by $\mathcal{I}_1$, $\mathcal{I}_2$) for upsampling, each time by a factor of 2, to get back to the original resolution. Finally, a mixing convolution module $C_m$ fuses the multichannel information, and the output is added to $\mathbf{h}_i$ following a residual connection to arrive at $\mathbf{h}_{i+1}$. The updated hidden representation $\mathbf{h}_{i+1}$ is fed into the next block. }
    \label{fig:net}
\end{figure}

Figure \ref{fig:net} depicts a block with $L=2$. \ivnet takes the vesicle configuration $\XX$ as input $\mathbf{h}_0$, and transforms it iteratively with multiple blocks to approximate a target output $Y$. The NN is trained by minimizing the relative error loss function $\mathcal{L} = \frac{\|\hat{Y} - Y\|_2}{\|Y\|_2} +\beta\frac{\|d\hat{Y}- dY\|_2}{\|dY\|_2}  $, where $Y$ is the ground truth, $\hat{Y}$  the network prediction, $dY$ and $d\hat{Y}$ their derivatives with respect to discretization. The second term is a relative error of $dY$ and is used to enhance smoothness of the prediction. The strength of the first term and second term can be tuned by paramter $\beta$. We use \texttt{AdamW} optimizer to train the networks with an initial learning rate of 0.001. 

Table \ref{tab:networks} summarizes the networks for the maps approximated by NNs in \vesnet, and the number of trainable parameters for them for $N=32$.

\begin{table}
\begin{center}
\begin{tabular}{cccc} \hline
 \bf Definition & \bf Map & \bf Network Size\\ \hline
         Relaxation flow map & $X_{\XX}: \XX \rightarrow \Delta \XX$ & 0.6M\\
         $\XX$ to advection term \eqref{eq:advectionMap} & $X^k_{\uu}: \XX \rightarrow \left(1-\calM\right) e^{ik\alpha}$  & 7.5M\\
         $\XX$ to self-tension term & $S_{\sigma}: \XX \rightarrow (\calL)^{-1}\left(\calP \calB \right)\XX$  & 0.3M\\
         $\XX$ to velocity-related tension term \eqref{eq:advTenMap} & $S^k_{\uu}: \XX \rightarrow (\calL)^{-1}\calP e^{ik\alpha}$  & 11.9M\\
         $\XX$ to SLI of Fourier basis in near-field \eqref{eq:NearFieldMap} & $N_{\XX}^k: \XX \rightarrow \calS[\XX, e^{ik\alpha}]\xx$  & 26.3M (5 layers)\\
         \hline
\end{tabular}
\caption{List of the networks for approximating the nonlinear maps. The networks acting on the Fourier basis functions are much larger as they merge $N_f$ subnetworks together.}\label{tab:networks}
\end{center}
\end{table}

As discussed in \secref{sec:nnoperators}, to handle arbitrary velocities or forces, we employ a decomposition strategy using the Fourier transform, allowing the network to learn on individual Fourier modes. This approach involves training $N_f$
subnetworks separately, each dedicated to a single Fourier mode. To fully utilize GPUs during inference, we merge all convolution and deconvolution modules of the $N_f$ subnetworks along the channel dimension, resulting in a larger and more parallelizable network. Since each subnet shares the same architecture, grouped convolutions can be used to parallelize operations while preserving the independence of the subnets within the merged network. In practice, the subnetworks are trained separately, and their weights are subsequently loaded into the merged network for inference. Although training different Fourier modes separately may, in theory, overlook cross correlation effects, our experiments show that training all modes altogether with one very large network does not increase the accuracy. Moreover, training the subnets individually requires less memory and has a smaller computational overhead. Therefore, we adopt a two-step approach: first training the subnets for different Fourier modes separately, and then merging them for inference.

\section{Algorithm and implementation}\label{sec:vesnet}
\begin{figure}
    \centering
    \includegraphics[width=0.95\textwidth]{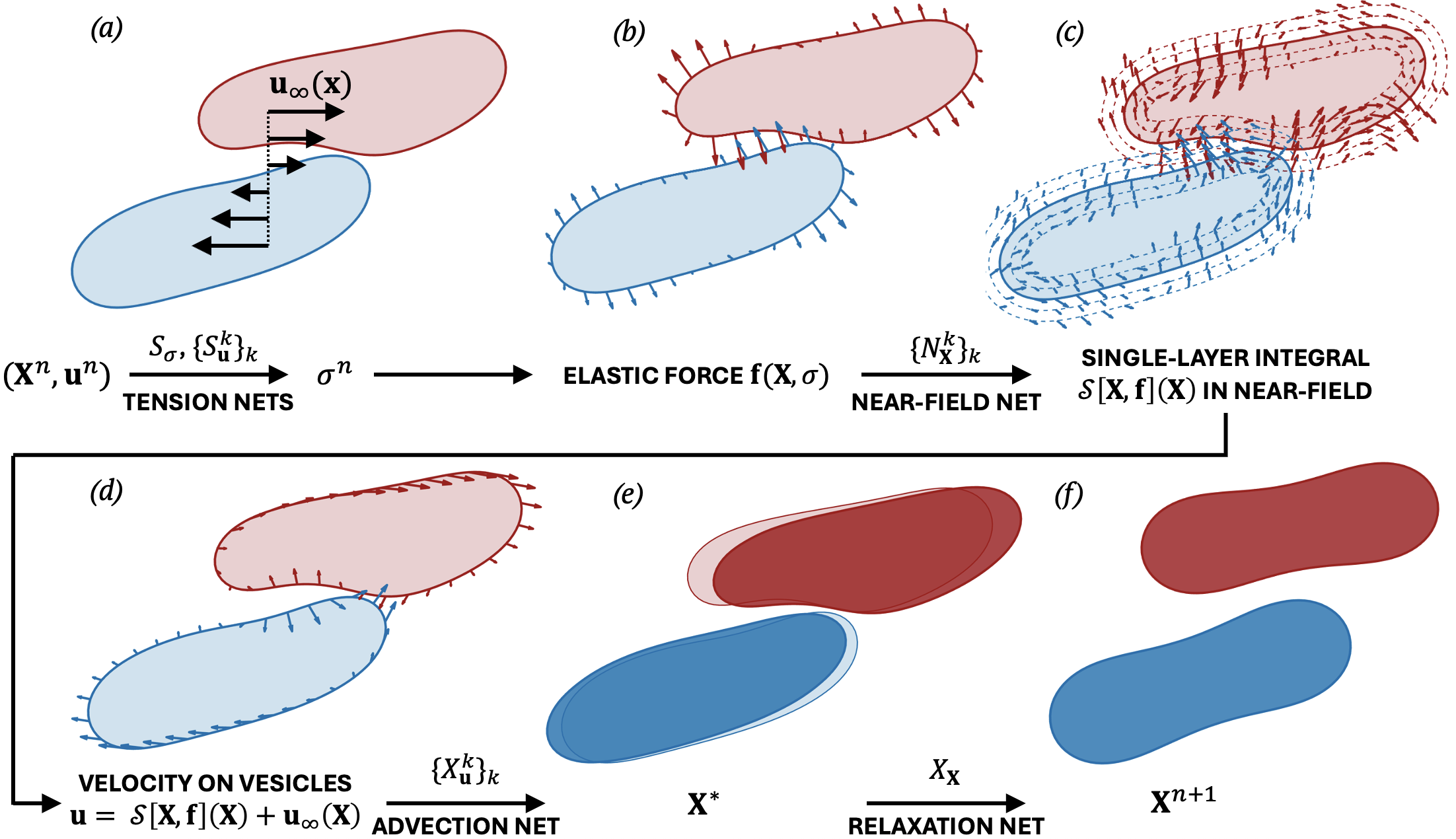}
    \caption{Main components of \vesnet. (a) Given the vesicle configurations at the current step $\XX^n$ suspended in shear flow and the velocity on each vesicle $\uu^n$, we find the tension $\sigma^n$ using the self-tension and the tension-velocity nets as approximations to $S_{\sigma}$ and ${S_{\uu}^k}_k$. (b) We compute the elastic membrane force $\ff = -\kappa_b\XX_{ssss}+(\sigma \XX_s)_s$ directly using $\XX^n$ and $\sigma^n$. The arrows on the vesicles show the force $\ff$. (c) We evaluate the single-layer integral of the force $\ff$ using the direct evaluation for the far-field, while applying the near-field network to correct the integration errors in the near-field. The dashed lines show the layers around each vesicle used for the interpolation of the single-layer integral in the near-field. (d) The arrows on the vesicle show the total velocity $\uu$ on the vesicles. (e) We update the vesicle configurations by first solving the advection problem using the advection net and the velocity found in (d). The dark colors show the updated vesicle configurations $\XX^*$ while the light colors show the vesicles at the current step. (f) Finally, the advected vesicles are relaxed to their next-step configurations $\XX^{n+1}$ through the relaxation net. The relaxation step here is exaggarated for visibility.}
    \label{fig:allNets}
\end{figure}

In this section, we discuss the GPU implementation of \vesnet to maximize parallelism and provide details on two important processing steps: arclength redistribution, and area and length correction. Additionally, we include a table comparing the computational time of \vesnet with that of BIEM. We present pseudocode for the time-stepping algorithm and its core components in \secref{sec:algos}.

\subsection{GPU implementation}
We utilize GPU's parallel computing capability and the PyTorch library to avoid using for-loops in \vesnet. Parallelism is exploited at two key levels. First, we discuss the parallelism at the tensor element level. Several operations in input (de-)standardization, including rotation, translation (line 2 in Algorithms \ref{alg:advect} and \ref{alg:tension}, line 8 in Algorithms \ref{alg:advect} and line 12 in Algorithm \ref{alg:singlelayer}) require all the coordinates of a vesicle go through affine transformations. We efficiently handle these transformations using vectorized tensor operators of PyTorch. Moreover, PyTorch’s broadcasting mechanism allows operands with different dimensions to be combined without copying data in memory, further improving efficiency. Second, the parallelism over the number of vesicles is crucial for scaling the system size. We achieve this through batching, enabling the networks to process multiple vesicles simultaneously. Additionally, we use batched Cholesky decomposition and triangular solves when computing RBF coefficients, and batched eigen decomposition to determine the principal axes of vesicles. We also exploit implicit batching in high-dimensional tensor operations, such as upsampling multiple vesicles concurrently via FFT.

Another technique to eliminate for-loops is advanced indexing, in which one tensor is used to index another, enabling selective access to specific elements or regions. We use integer index tensors to reorder the coordinates when (de-)standardizing the vesicle configurations. In computing the single layer integral \eqref{eq:single_layer} between vesicles, we perform direct integration with boolean masks to set neighboring vesicles' contribution to zero, therefore avoiding the need to subtract it later. We then use the same masks to identify the entries that require near-singular integration.

\subsection{Redistributing points along arclength}
\label{sec:redist_arclength}
As part of the standardization step, we redistribute membrane points to ensure they are approximately equispaced in arclength. The procedure is the following. Given a vesicle that may not have equispaced points, we compute arclength $\{l_k\}_{k=1}^N$ at $\{\alpha_k = 2k\pi/N \}_{k=1}^N$, build a piece-wise linear interpolant, and then infer corresponding 
$\{\tilde{\alpha}_k\}_{k=1}^N$ for the equispaced arclength $\{\tilde{l}_k = k/N \}_{k=1}^N$. In practice, since PyTorch does not have linear interpolation routines, we first do a search in the increasing arclength sequence to locate the interval containing the target arclength, then interpolate based on the two endpoints of the interval. We also batch this operation. 

Since the relationship between $\alpha_k$ and arclength $l_k$ is not linear, simple linear interpolation may introduce additional errors. To mitigate this issue, we first upsample the membrane points by a factor of $6$ (heuristically chosen) before other steps. Empirically, this approach yields accurate and stable simulations and is faster and simpler than using nonlinear interpolation schemes such as cubic splines. The computational complexity is dominated by the FFT in computing arclength, as well as searching within the monotonically increasing arclength sequence $N$ times, both of which scale as $O(N\log{N})$ per vesicle.

\subsection{Correction of errors in area and length}
Due to the inextensibility and the incompressibility constraints, the arclength and the area enclosed by the membrane are conserved quantities. However, the accumulation of numerical errors leads them to drift from their initial values \citep{kabacaoglu-biros-e18}. In order to avoid the drift and maintain stability of simulation for long horizons, we correct these errors. Suppose that a vesicle initially has area $A_0$ and length $L_0$, and that $\mathbf{X}^n$ is the configuration at time $n$, we consider the following constrained optimization problem to obtain a new configuration $\tilde{\mathbf{X}}^n$, where the constraints require the vesicle’s area and length to remain the same as initial values:
\begin{equation*}
\operatorname{argmin}\|\tilde{\mathbf{X}}^n-\mathbf{X}^n\|^2, \quad\text{such that } A_n=A_0, 
 L_n=L_0.\\
\end{equation*}
For fast solution of this constrained optimization problem, we formulate an unconstrained objective function following the augmented Lagrangian method:
\begin{equation*}
\operatorname{min}\|\tilde{\mathbf{X}}^n-\mathbf{X}^n\|^2  +\frac{\mu^k}{2} \sum_{i \in \mathcal{E}} c_i(\mathbf{X})^2 - \sum_{i \in \mathcal{E}} \lambda_i^k c_i(\mathbf{X}),
\end{equation*}
where $c_i(\mathbf{X})$ denotes constraints of area and arclength, and $\mathcal{E}$ denotes the indices for equality constraints. This objective contains both quadratic penalty terms and Lagrangian terms, and improves upon classical Lagrange multiplier method. 

Starting with initial $\mu^0$ and $\lambda_i^0$, the objective is reduced by running a few gradient-based minimization steps, and the Lagrange multipliers are updated according to the formula
\begin{equation*}
\lambda_i^{k+1}=\lambda_i^k-c_i\left(x_k\right) / \mu_k, \quad \text { for all } i \in \mathcal{E} \text {. }
\end{equation*}
Then we also reduce the penalty parameter by choosing $\mu^{k+1} \in (0, \mu^k)$, and, with the new parameters, minimize the objective again and start another iteration of the procedure. 

For GPU implementation, we treat vesicle configurations as trainable parameters and employ the \texttt{Adam} optimizer to minimize the objective given $\mu^k$ and $\{\lambda_i^{k}, i \in \mathcal{E}\}$. This approach leverages the strengths of GPU acceleration and utilizes a well-established optimizer to enhance computational efficiency. In practice, we perform area and length corrections only when the relative errors exceed $1\%$, thereby further amortizing the computational cost over multiple time steps.

\subsection{Algorithms}\label{sec:algos}
Here, we summarize one time step of \vesnet for multiple vesicles (also see \figref{fig:allNets}). The input consists of the configurations of $M$ vesicles, $\XX$, discretized with $N$ points, the material parameters (bending stiffness $\kappa_b$ and viscosity $\mu$), the background velocity $\uback$ and the time step size $\Delta t$. The main algorithms are presented below: time stepping, computing tension, solving the advection problem and computing the near-singular integrals with NNs. We refer the reader to \citet{kabacaoglu-biros19pre} for the details of the (de-)standardization algorithms. 

\begin{algorithm}[!htb]
	\caption{\vesnet: Time stepping scheme for multiple vesicles} 
        \label{alg:time_step}
	\begin{algorithmic}
            \STATE{Given $\XX^n$ and $\sigma^n$}
            \STATE $\uu^n = \mathtt{ComputeSingleLayerIntegral} (\XX^n, \sigma^n) + \uback(\XX^n)$ // Algorithm \ref{alg:singlelayer}
            \STATE $\XX^* = \mathtt{SolveAdvection}(\XX^n,\uu^n)$ Algorithm \ref{alg:advect}
            \STATE $\XX^{n+1} = \mathtt{ApproximateX_{\XX}}(\XX^*)$ 
            \STATE $\sigma^{n+1} = \mathtt{ComputeTension}(\XX^{n+1}, \ubar^n)$ // Algorithm \ref{alg:tension}
            \STATE $\XX^{n+1} = \mathtt{ApplyCorrectionAlgorithms}(\XX^{n+1},\mathtt{area},\mathtt{length})$
            \RETURN $\XX^{n+1}$ and $\sigma^{n+1}$
	\end{algorithmic} 
\end{algorithm}

\begin{algorithm}[!htb]
	\caption{$\mathtt{ComputeTension}$} 
        \label{alg:tension}
	\begin{algorithmic}[1]
        \STATE // Given $\XX$ and velocity induced on the vesicle $\uu$
        \STATE $(\XX^o, \tau, \theta, \mathtt{order}) = \mathtt{standardize}(\XX)$
        \STATE // Here, the standardization parameters are found, translation $\tau$ to center, rotation $\theta$ to align, and point ordering
        \STATE $\sigma_{\mathrm{int}}^o = \mathtt{ApproximateS_{\sigma}}(\XX^o)$
        \STATE $\uu^o = \mathtt{rotateAndOrder}(\uu,\theta,\mathtt{order})$
        \STATE $\{\widehat{\uu}^o_k \}_{k = 1}^{N_f} = \mathtt{fft}(\uu^o)$
        \STATE $\{ V_k \}_{k = 1}^{N_f} = \mathtt{ApproximateS_{\uu}^k}(\XX^o)$
        \STATE $\sigma_{\mathrm{ext}}^o = \sum_k^{N_f} V_k \widehat{\uu}^o_k$
        \STATE $\sigma^o = \sigma_{\mathrm{int}}^o + \sigma_{\mathrm{ext}}^o$
        \STATE $\sigma = \mathtt{orderBack}(\sigma^o,\mathtt{order})$
        \RETURN $\sigma$.
	\end{algorithmic} 
\end{algorithm}

\begin{algorithm}[!htb]
	\caption{$\mathtt{SolveAdvection}$} 
        \label{alg:advect}
	\begin{algorithmic}[1]
        \STATE // Given $\XX$ and velocity induced on the vesicle $\uu$
        \STATE $(\XX^o, \tau, \theta, \mathtt{order}) = \mathtt{standardize}(\XX)$
        \STATE $\uu^o = \mathtt{rotateAndOrder}(\uu, \theta, \mathtt{order})$
        \STATE $\{ \widehat{\uu}_k^o\}_{k=1}^{N_f} = \mathtt{fft}(\uu^o)$
        \STATE $\{V_k \}_{k=1}^{N_f} = \mathtt{ApproximateX_{\uu}^k}(\XX^o)$
        \STATE $(1-\calM^o)\uu^o = \sum_{k}^{N_f} V_k \uu^o_k$
        \STATE $\XX^{*,o} = \XX^o + \Delta t(1-\calM^o)\uu^o$
        \STATE $\XX^* = \mathtt{destandardize}(\XX^{*,o},\tau,\theta,\mathtt{order})$
        \RETURN $\XX^{*}$
	\end{algorithmic} 
\end{algorithm}

\begin{algorithm}[!htb]
	\caption{$\mathtt{ComputeSingleLayerIntegral}$} 
        \label{alg:singlelayer}
	\begin{algorithmic}[1]
		\STATE // Given $\XX$ and their membrane forces $\ff$ 
            \STATE Initialize $\uu_{\mathrm{near}} = 0$
            \STATE $\uu_{\mathrm{far}} = \mathtt{DirectSingleLayerIntegral}(\XX, \ff)$
             
            \STATE $\XX_q = \mathtt{FindNearFieldPoints}(\XX)$
             
            \IF{There is any $\XX_q$}
            \STATE $(\XX^o, \tau, \theta, \mathtt{order}) = \mathtt{standardize}(\XX)$
            \STATE $\xx^o_{\mathrm{layers}} = \mathtt{CreateNearLayers}(\XX^o)$
            \STATE $\{V_k\}_{k=1}^{N_f} = \mathtt{ApproximateN_{\XX}^k}(\XX^o)$
            \STATE $\ff^o = \mathtt{rotateAndOrder}(\ff,\theta,\mathtt{order})$
            \STATE $\{ \widehat{\ff}_k\} = \mathtt{fft}(\ff^o)$
            \STATE $\uu_{\mathrm{layers}}^o = \sum_k^{N_f} V_k \widehat{\ff}_k$
            \STATE $\xx_{\mathrm{layers}} = \mathtt{destandardize}(\xx_{\mathrm{layers}}^o, \tau, \theta, \mathtt{order})$
            \STATE $\uu_{\mathrm{layers}} = \mathtt{rotateAndOrderBack}(\uu_{\mathrm{layers}}^o,\theta,\mathtt{order})$
            \STATE $\uu_{\mathrm{near}} = \mathtt{interpolateInNearField}(\xx_{\mathrm{layers}},\uu_{\mathrm{layers}}, \XX_q)$
            \ENDIF
        \STATE $\uu = \uu_{\mathrm{far}} + \uu_{\mathrm{near}}$
        \RETURN $\uu$      
	\end{algorithmic} 
\end{algorithm}

\subsection{Timings}
In Table \ref{tab:timings} we show the inference times for the main algorithms in one time step and show their scaling over the numbers of vesicles $M$. \vesnet and BIEM measurements are made on the NVIDIA GH200 Grace Hopper Superchip. Direct evaluation of the single layer integral \eqref{eq:single_layer} is performed for vesicle-vesicle interactions in the far-field, and is performed twice (denoted by $\times 2$) since the traction jump $\ff$ was updated (see input of Algorithm \ref{alg:singlelayer}). This computation is done without any approximation, and scales quadratically with $M$ as validated in the table. Applying the fast multipole method \citep{greengard-rokhlin87} will significantly reduces the cost and make it scale as $O(M)$, but an applicable GPU implementation is beyond the scope of this paper. The inference times for the networks are small compared to the direct evaluation of the single layer integral, and scales linearly or sublinearly with $M$. Particularly, for the relaxation and self-tension networks, the cost remains nearly constant across different $M$ values, due to their small sizes (see Table \ref{tab:networks}). Another notable part is redistributing points along arclength (see \secref{sec:redist_arclength}). This algorithm scales linearly with respect to $M$, and is called three times in a time step, which are standardizing $\XX^n$ and $\XX^* $, and adjusting arclength of $\XX^{n+1}$.

\begin{table}
\begin{center}
\begin{tabular}{lcccc} \hline
\diagbox[width=24em]{Name}{Time}{$M$} & 48 & 504 & 1020 & 2000\\\hline
    Network approximation for near-singular integral  & 0.05 & 0.10 & 0.18 & 0.34\\
    Direct evaluation of single layer integral ($\times 2$) & 0.00 & 0.05 & 0.17 & 0.66 \\
    Velocity-related tension solve & 0.02 & 0.04 & 0.07 & 0.13\\
    Self-tension solve & 0.01 & 0.01 & 0.01 & 0.02 \\
    Advection solve & 0.02 & 0.04 & 0.08 & 0.15 \\
    Relaxation solve & 0.01 & 0.01 & 0.01 & 0.02 \\
    Point distribution equally in arclength ($\times 3$) & 0.03 & 0.03 & 0.03 & 0.04\\
    \hline
    Total (\vesnet) & 0.14 & 0.29 & 0.56 & 1.37\\
    \hline    
    Total (BIEM $N$=32) & 0.53 & 1.44 & 3.68 & 12.88\\
    \hline    
\end{tabular}
\caption{Timings (unit: second) for the main algorithms in one time step with \vesnet. The Total (\vesnet) row shows total time needed for not only the main algorithms listed in the table but also other processing steps (whose execute times are negligible). The total time needed for BIEM with $N$=32 is shown in the last row for comparison. All computations are done in single precision (\texttt{float32}). For a fair comparison, \vesnet and BIEM both use RBF interpolation for near singular-integration. For BIEM, the input layers of velocities for RBF are computed exactly.
%
}\label{tab:timings}
\end{center}
\end{table}


Although the networks have millions of weights, \vesnet is faster than BIEM. The main reason is the following. In every time step, BIEM solves a preconditioned linear system of size $O(M)$ through GMRES (see \secref{sec:biem-time-discretization}), and each iteration of GMRES requires computing vesicle-vesicle interactions by evaluating the single-layer integral on iterate $\XX$. Evaluating such integral is very expensive compared to network inference, as evidenced by Table \ref{tab:timings}. Even though the networks have lots of weights, these weights are mostly used for matrix multiplications under the hood, which is naturally parallelizable on GPU and utilizes highly optimized routines of PyTorch.

\section{Results}\label{sec:results}

In this section, we discuss the results of several numerical experiments to assess the accuracy and speed of \vesnet. 
\begin{enumerate}
    \item \textit{Relaxation (\secref{sec:relax}):} We simulate the relaxation of a deformed vesicle which is not in the dataset used to train the networks to its equilibrium configuration. The result shows that (i) the relaxation net accurately captures the equilibrium and the membrane bending energy, and (ii) it generalizes to unseen vesicle configuration.  
    \item \textit{Pouiseille flow (\secref{sec:pois}):} We simulate lateral migration and deformation of a vesicle in a parabolic distribution of the velocity field (e.g., Poiseuille flow) where both the advection and the relaxation nets are used. The results show that \vesnet accurately finds the equilibrium lateral positions and configurations of vesicles, even under large shear gradients and high degrees of confinement.
    \item \textit{Shear flow (\secref{sec:shear}):} We simulate two vesicles in shear flow where all the nets are used. The results demonstrate that the overall accuracy of \vesnet and that the proposed near-singular integration scheme is essential for stability and accuracy of the simulations.
    \item \textit{Dense suspensions (\secref{sec:denseSuspension}):} We simulate dense vesicle suspensions in Taylor-Green flow and in Pouiseille flow to show the \vesnet's scaling and the accuracy in certain suspension statistics. \vesnet demonstrates superior accuracy in quantifying dense suspension dynamics compared to BIEM at equivalent spatio-temporal resolutions.
\end{enumerate}

\subsection{Simulation parameters} We compare \vesnet  against the ground truth obtained with our in-house BIEM code \citep{kabacaoglu-biros-e18}. We consider vesicles that have the same interior fluid viscosity as the exterior fluid viscosity and have the reduced area 0.65. In the ground truth simulations, vesicles are discretized with $N = 128$ points. The ground truth simulations are labeled as BIEM. We explain the problem-specific details below. 

\begin{figure}
    \centering
    \includegraphics[width=0.98\textwidth]{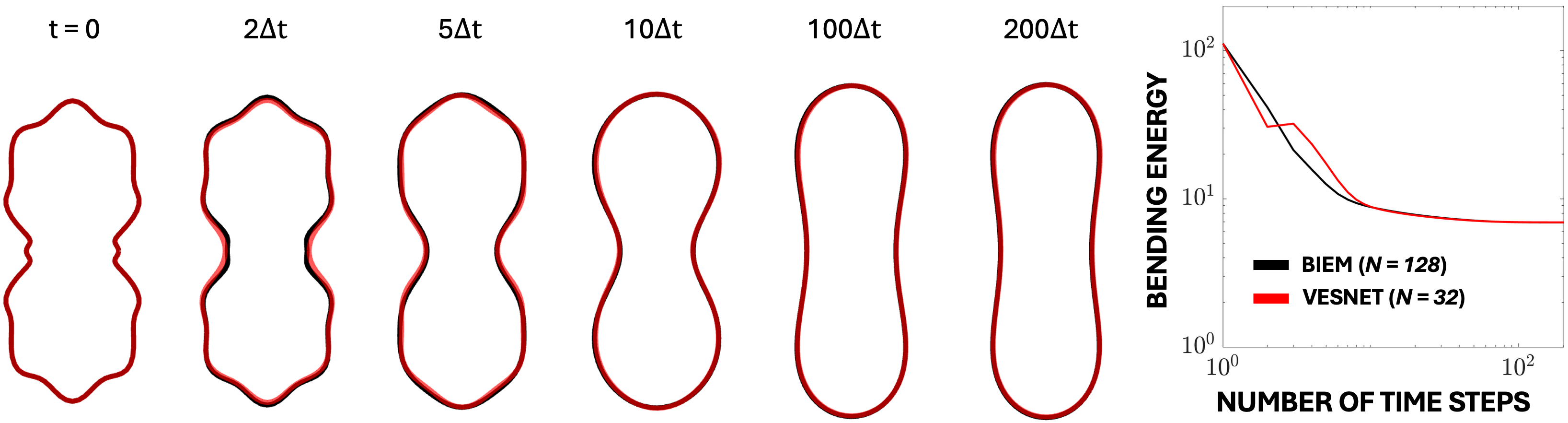}
    \caption{Snapshots of an arbitrary shaped vesicle in a stationary fluid. Vesicle relaxes to an equilibrium shape which minimizes the membrane bending energy (also shown on the right). We discretize vesicle with $N = 128$ points and use the same time step size $\Delta t = 10^{-5}$ in BIEM and \vesnet.}
    \label{fig:FigRelax}
\end{figure}

\subsection{Relaxation}\label{sec:relax}
We consider a deformed vesicle suspended in a stationary fluid. The vesicle reaches an equilibrium shape which minimizes the vesicle bending energy \citep{seifert91}. Due to the local inextensibility constraint, the equilibrium configuration can be different than a circle which is the equilibrium for extensible vesicles. \figref{fig:FigRelax} shows the relaxation of an arbitrary shaped vesicle. Since there is only one vesicle and the fluid is stationary, only \eqref{eq:relaxation_disc} is solved during the simulations. We also plot the bending energy of the membrane, $E_b = \kappa_b / 2 \int_{\gamma} c^2 ds$ where $c$ is the local curvature. The results show that although the initial configuration is outside the dataset, \vesnet accurately finds the equilibrium shape and the bending energy. 

\subsection{Poiseuille flow}\label{sec:pois}

\begin{figure}
    \centering
    \includegraphics[width=0.98\textwidth]{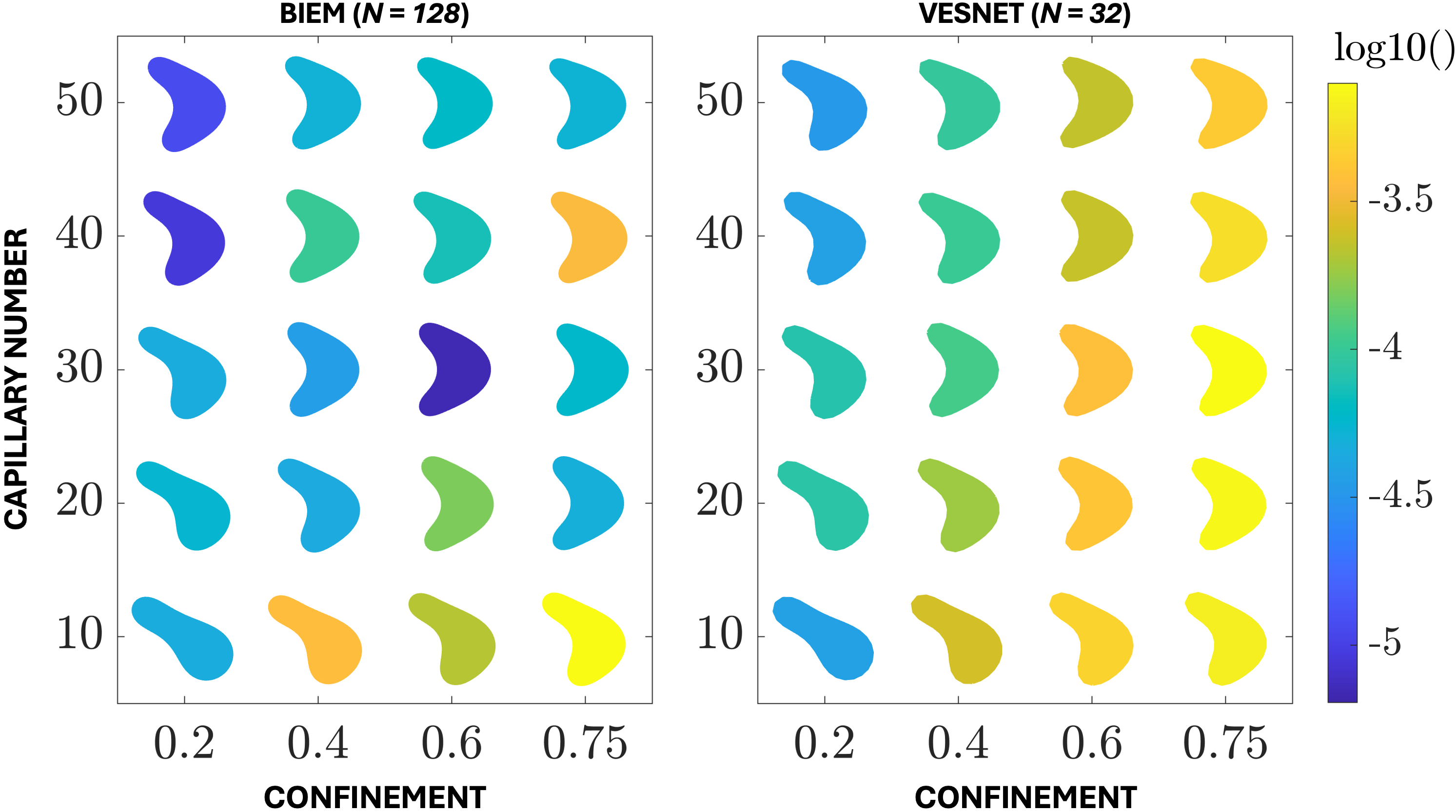}
    \caption{Phase diagram of equilibrium vesicle shapes in Poiseuille flow. BIEM (on the left) and \vesnet (on the right). The colorbar shows the vertical velocity in equilibrium normalized with the net velocity in equilibrium. See \figref{fig:FigLateral} for the equilibrium lateral positions.}
    \label{fig:FigPhase128}
\end{figure}

\begin{figure}
    \centering
    \includegraphics[width=0.98\textwidth]{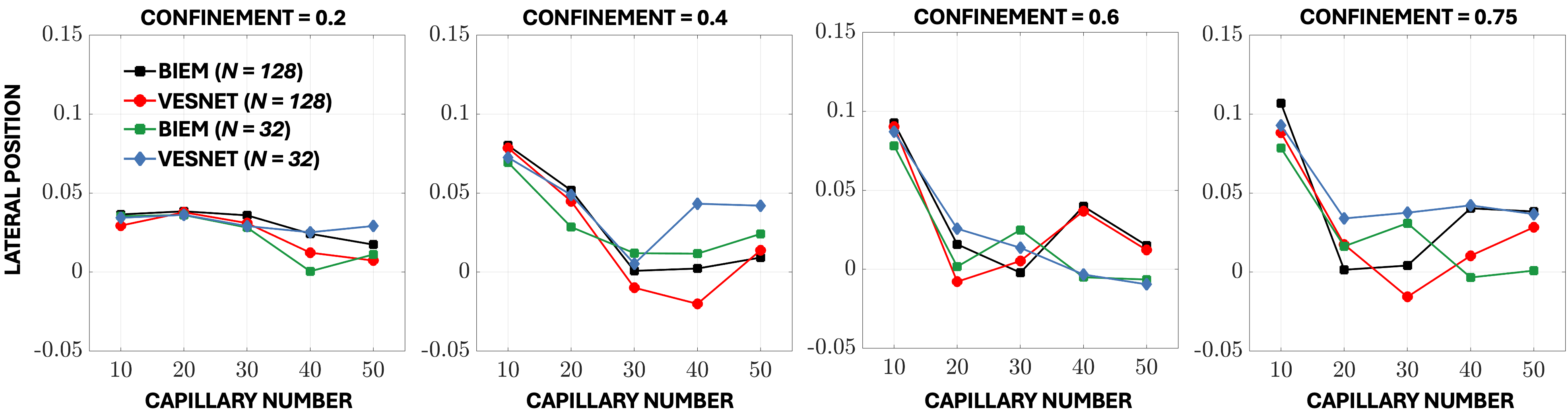}
    \caption{Equilibrium lateral position in Poiseuille flow. Each figure corresponds to different degree of confinement Cn. The lateral position is normalized with $W$. See \figref{fig:FigPhase128} for the equilibrium vesicle configurations. In addition to the simulations shown in \figref{fig:FigPhase128} we perform BIEM simulations with $N = 32$ and \vesnet simulations with $N = 128$ to show the differences due to the spatial resolution.}
    \label{fig:FigLateral}
\end{figure}

We consider a vesicle in a Poiseuille flow, i.e., the imposed flow is $\uback(y) = u_{\max}[1-(y/W)^2]\mathbf{e}_x$. Here, $u_{\max}$ determine the flow strength and $y = \pm W$ is where the imaginary walls are. This flow profile has a flow line curvature $2u_{\max}/W^2$ \citep{kaoui-misbah09}. In this flow, a vesicle that is initially off-centered migrate towards the low shear rate region. That results in the formation of a cell-free layer near the vessel walls in microcirculation (i.e., the Fahraeus-Lindqvist effect) and consequently the decrease of apparent blood viscosity \citep{fahraeus-lindqvist31,popel-johnson05,kumar-graham12}. The vesicle eventually yields an equilibrium configuration and lateral position. Length is measured in units of the vesicle radius $R$ defined as the radius of a circle having the same enclosed area. In all simulations, vesicle is initialized at $y = W/2$ with its principal axis is parallel to the $x$-axis. We investigate equilibrium vesicle dynamics in this flow using two dimensionless parameters: the degree of confinement Cn and the capillary number Ca
\begin{equation*}\label{eq:dimensionlessParams}
    \text{Cn} = \frac{R}{W} \quad \text{and} \quad \text{Ca} = \frac{2\eta u_{\max}R^3}{\kappa_b W}.
\end{equation*}
Ca is $\bigO(1)$ for healthy red blood cell flow in a vessel \citep{fung90}. We define Ca different in this work compared to our prior work \citep{kabacaoglu-biros19pre} where the degree of confinement is fixed to $R/W = 0.1$. Note that one-fifth of the Ca in \citet{kabacaoglu-biros19pre} corresponds to the Ca in the present work.  

We perform simulations for Cn $ = (0.2, 0.4, 0.6, 0.75)$ and Ca $ = (10, 20, 30, 40, 50)$ until vesicles reach equilibrium with BIEM and \vesnet using $N = 128$ and $N = 32$ points to discretize a vesicle, respectively. The stable largest time step size in the simulations is $10^{-5}$ for Ca $ \leq 30$ and $5 \times 10^{-6}$ for Ca $ \geq 40$. Figure \ref{fig:FigPhase128} shows the phase diagram of the equilibrium configurations where vesicles are colored based on their vertical velocity in equilibrium which is expected to be zero as the lateral position does not change in the equilibrium. Figure \ref{fig:FigLateral} shows the vesicles' equilibrium lateral positions. 

The degree of vesicle deformation and the curvature of the flow increase with higher capillary numbers and greater confinement, making the simulation of vesicle dynamics increasingly challenging. \vesnet is more accurate for low capillary number and degree of confinement than it is for their higher values. BIEM and \vesnet give qualitatively similar equilibrium configurations. However, there are quantitative differences in the equilibrium lateral positions. In our prior work \citep{kabacaoglu-biros19pre}, the confinement was Cn $ = 0.1$ and the highest capillary number that can be stably resolved was 45 (calculated based on the Ca definition here). In this work, we improved the new solver's capabilities by optimizing the neural network hyperparameters and architecture.

We investigate how the number of discretization points $N$ affect the accuracy and the stability of \vesnet in this example. Thus, we repeat the simulations with \vesnet using $N = 128$, and BIEM with $N = 32$. The results show that \vesnet has accuracy similar to or better than BIEM at the spatial resolution of $N = 32$ points. \vesnet's accuracy improves as the resolution $N$ increases.

\begin{figure}
    \centering
    \includegraphics[width=0.98\textwidth]{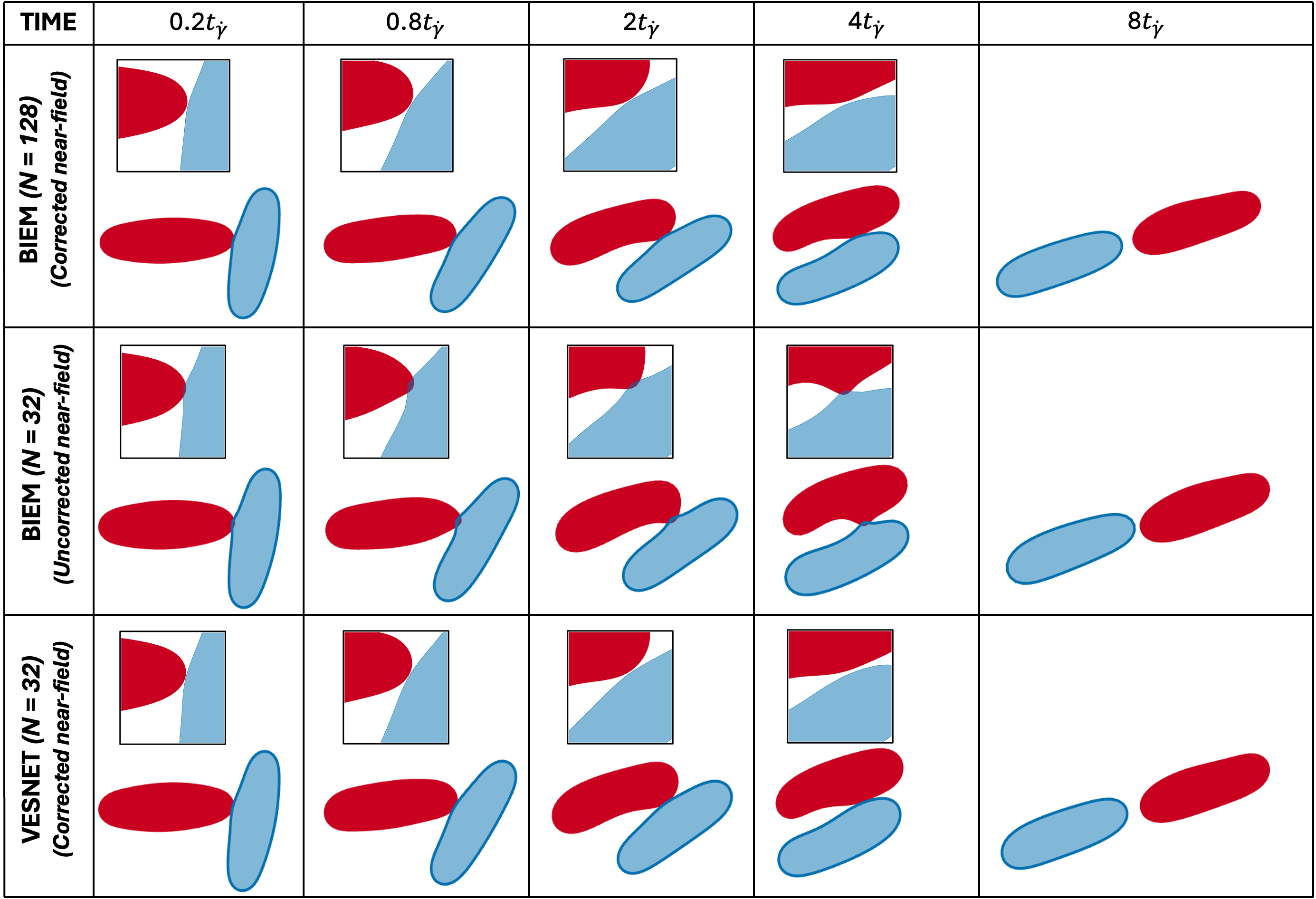}
    \caption{Two vesicles in shear flow. Each row shows snapshots from simulations. The time is nondimensionalized by the time scale associated to the vesicle deformation due to viscous forces $t_{\dot{\gamma}}$. The ground truth in the first row is obtained with BIEM using $N = 128$ points and the near-singular integration scheme \citep{ying-zorin-e06}. The middle row shows the simulation performed with BIEM where 32 points are used to discretize a vesicle and the near-field interactions are not corrected. In this simulation, the vesicles collide at $0.8t_{\dot{\gamma}}$. The last row shows the simulation performed with \vesnet where 32 points are used to discretize a vesicle and the near-field is corrected with the proposed near-field correction scheme. \vesnet accurately handles close vesicle interactions and avoids collision.}
    \label{fig:FigShearSnaps}
\end{figure}

\begin{figure}
    \centering
    \includegraphics[width=0.8\textwidth]{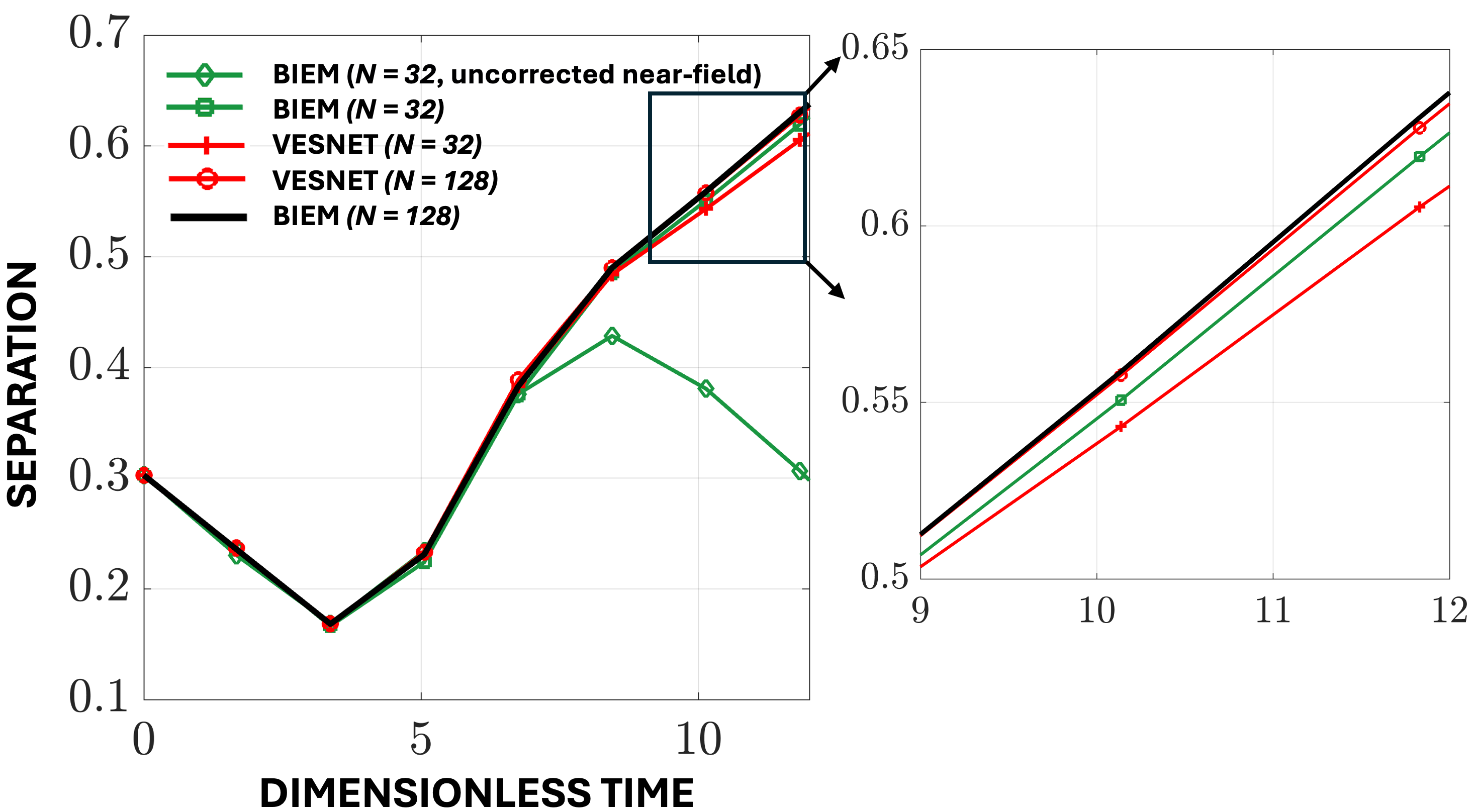}
    \caption{Separation between two vesicles in shear flow. In the simulations, the near-singular integrals are corrected except the one with BIEM using $N = 32$. \vesnet accurately captures the distance between two vesicles after close interaction with $N = 32$ points. The accuracy improves when the resolution increases. If the near-field is corrected, BIEM is more accurate than \vesnet at $N = 32$.}
    \label{fig:FigShearSep}
\end{figure}

\subsection{Shear flow}\label{sec:shear}
We consider two vesicles in an unbounded shear flow $\uback = \dot{\gamma}(y,0)$ where $\dot{\gamma}$ is the shear rate. In this example, we aim at showing the necessity of using a specialized quadrature for near-singular integrals to stably and accurately simulate vesicles in close contact.

There are two time scales related to vesicle dynamics in shear flow. One of them is associated with the vesicle deformation due to viscous forces, i.e., $t_{\dot{\gamma}} = 1/\dot{\gamma}$; and the other one is about the membrane's bending response that works to bring the vesicle back to the preferred curvature, i.e., $t_{\kappa} = \eta R^3/\kappa_b$ \citep{vlahovska-gracia07}. Here, we define the time in the units of $t_{\dot{\gamma}}$. The capillary number is the ratio of these two time scales, Ca $ = \eta R^3 \dot{\gamma}/\kappa_b$.  The time horizon is 20$t_{\dot{\gamma}}$. We perform our simulations at $Ca \approx 5$ using BIEM and \vesnet with $N = (32, 128)$ points. The largest stable time step size is 0.01$t_{\dot{\gamma}}$ and 0.02$t_{\dot{\gamma}}$ for $N = 32$ and $N = 128$, respectively (regardless of BIEM or \vesnet). In the BIEM simulations, we use the near-singular integration scheme proposed in \citet{ying-zorin-e06}. The vesicles are initialized such that they are almost contacting. One vesicle is at $(x, y) = (-2.3R,0.3R)$ with its principal axis parallel to the $x$-axis and the other one is at $(x, y) = (0, 0)$ with its principal axis perpendicular to the $x$-axis. Hence, the vesicle on the left moves towards the other vesicle which tilts due to the imposed shear flow. This is shown in the top row in \figref{fig:FigShearSnaps} where the snapshots from the ground truth simulation performed with BIEM using $N= 128$ points and the near-singular integration scheme are presented.

When we use BIEM to perform the simulations, we do not observe any collision between vesicles at both resolutions $N = 32$ and $N = 128$ if the near-singular integration scheme is used. If the scheme is not used, then vesicles collide when $N = 32$ (see the middle row in \figref{fig:FigShearSnaps}). BIEM with $N = 128$ can still handle the collision even when the near-singular integration scheme is not used since the resolution is fine enough to capture the single layer integral accurately. \vesnet leads the vesicles to collide for $N = (32, 128)$ points if the single layer integral is not corrected for the near-field points. We observe that the collision can be handled using the proposed near-field correction scheme. See the third row in \figref{fig:FigShearSnaps} for the snapshots. One of the quantities of interest in this example is the separation between the vesicles' centers shown in \figref{fig:FigShearSep}. Here, we observe that \vesnet with  $N = 128$ points and near-singular integral corrections most closely matches the ground truth. Among the low resolution simulations (i.e., $N = 32$), BIEM is more accurate than \vesnet. However, if we do not correct the near-singular integrals at this resolution, vesicle-vesicle collision leads to an erroneous result (see BIEM with $N = 32$ points and uncorrected near-field). 

These experiments clearly show that the near-singular integration scheme proposed in this work based on neural networks can effectively be used to correct the errors in nearly singular integrals. The proposed method has an advantage of not requiring any singular quadrature to evaluate the single layer integral on a vesicle itself which is an expensive computation in 3D \citep{malhotra-biros-e17}. 

\begin{figure}
    \centering
    \includegraphics[width=0.75\textwidth]{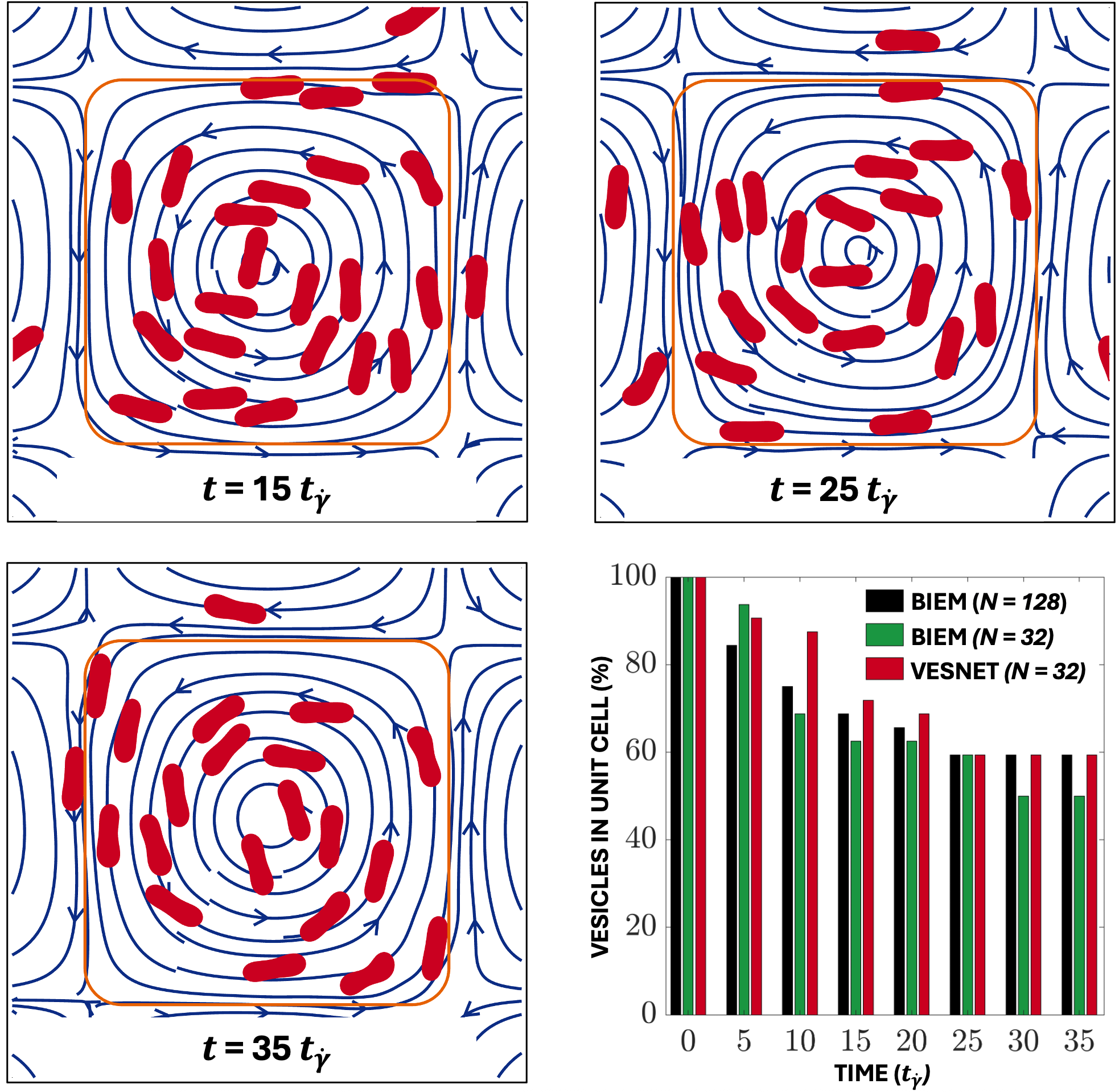}
    \caption{Dense suspension in Taylor-Green flow with velocity \eqref{eq:tayGreenFlow}. There are 32 vesicles in the unit cell initially. The time scale of the flow is defined based on the shear rate, i.e., $t_{\dot{\gamma}} = W/U$. We study vesicle migration from the unit Taylor-Green cell (highlighted with orange rectangle in the figures) due to the background flow and hydrodynamic interactions. Initially, the suspension occupies 25\% of the unit cell. We observe vesicle drainage from the cell (see top left for time $t = 15 t_{\dot{\gamma}}$, top right for time $t = 25 t_{\dot{\gamma}}$ and bottom left for time $t = 35 t_{\dot{\gamma}}$ in the ground truth simulation). We simulate this case with BIEM ($N = 128$), BIEM ($N = 32$) and \vesnet ($N = 32$). We compare the accuracy of the latter two based on the ratio of the vesicles remaining in the unit cell to the total number of vesicles (see bottom right). \vesnet accurately captures the ground truth due to more accurate approximations to the integro-differential operators than BIEM at a low resolution.}
    \label{fig:denseTGstats}
\end{figure}

\subsection{Dense suspensions}\label{sec:denseSuspension}

We, now, consider a more challenging experiment where we simulate dense suspensions with area fraction 25\% in Taylor-Green flow. The velocity field in Taylor-Green flow is 
\begin{equation}\label{eq:tayGreenFlow}
\uback = U\left[\sin(\pi x/W)\cos(\pi y/W), -\cos(\pi x/W)\sin(\pi y/W)\right],    
\end{equation}
where $U$ is the flow strength and $W$ is the length of the periodic cell of the flow. We define the time scale as $t_{\dot{\gamma}} = W/U$. In the first test, we initialize 32 vesicles that are initially pretty close to each other and occupying 25\% of the periodic cell. The ratio of the vesicles' arclength to the size of the unit cell $W$ is 0.4. We obtain a ground truth simulation using BIEM where we use $N = 128$ points to discretize a vesicle and a fixed time step size of $\Delta t =$ 1E-5. We, then, perform simulations with BIEM and \vesnet with $N = 32$ and $\Delta t = $ 1E-5. Vesicles show cross-streamline migration as in Poiseuille flow even in their dilute flows in Taylor-Green cell \citep{kabacaouglu2023cross}. Inter-vesicle interactions in their dense suspensions also contribute to the migration from the unit cell. We are interested in the statistics of vesicle migration, i.e., the percentage of vesicle remaining in the cell (see \figref{fig:denseTGstats}). The ground truth simulation ($N=128$) shows that approximately 60\% of the vesicles stay in the cell in statistical equilibrium. Simply reducing the resolution in BIEM to $N=32$ leads to overestimated migration. In the equilibrium (not shown) only 40\% of the vesicles remain in the cell. \vesnet reaches an equilibrium around the time $25 t_{\dot{\gamma}}$ and accurately captures the equilibrium density of the vesicles in the cell. 

To further demonstrate the computation time per time step and ability to resolve complex inter-vesicle interactions in dense suspensions, we simulate 2,000 vesicles in Poiseuille flow (see \secref{sec:pois}) with 30\% area fraction of vesicles. The dimensionless parameter values for this flow are Ca = 2 and Cn = 0.025. We show the snapshots from the simulation in \figref{fig:2000Parabolic}. We compare this simulation with the red blood cell flow in arterioles. In our simulations, a compute-length unit corresponds to 22 $\mu$m (the largest axis of red blood cell is 8 $\mu$m and 0.35 compute-length unit). To find a compute-time unit in terms of seconds, we consider the blood viscosity equal to the water viscosity (i.e., 0.001 Pa.s), the bending stiffness $\kappa_b = 10^{-19}$ J \citep{vlahovska-gracia07, kaoui-misbah-e09b}. The maximum velocity in the blood flow with Ca = 2 is 500 $\mu$m/s. Hence, one compute-time unit is 110s. One time step with \vesnet in this simulation takes 1.4s and corresponds to 0.001s in real time of blood flow.

\section{Conclusions}
In this study, we introduced \vesnet, a machine learning-accelerated solver designed to address the computational challenges inherent in simulating Stokesian particulate suspensions. \vesnet integrates the Iterated V-shape Net (\ivnet), with boundary integral equation method (BIEM) to significantly accelerate the solution of complex integro-differential equations governing vesicle dynamics. Our numerical experiments demonstrate that \vesnet not only achieves substantial computational speed-up to two orders of magnitude over traditional BIEM approaches—but also accurately captures critical flow characteristics, even for dense particulate suspensions and highly deformable vesicles. The NN-based approximation of near-singular integrals further enhances the solver’s stability and accuracy.  

Although \vesnet is currently limited to 2D simulations without viscosity contrasts, it provides a solid foundation for extension to 3D simulations and more complex biological and industrial systems. Future work includes addressing these limitations and developing efficient ML-based collision handling techniques. \vesnet thus represents a significant step towards real-time simulation capabilities, with broad implications for biomedical engineering, health monitoring devices, and advanced manufacturing processes.

\appendix
\section{Accuracy of near-singular integration algorithm}\label{sec:appendixNear}
\begin{figure}
    \centering
    \includegraphics[width=0.95\textwidth]{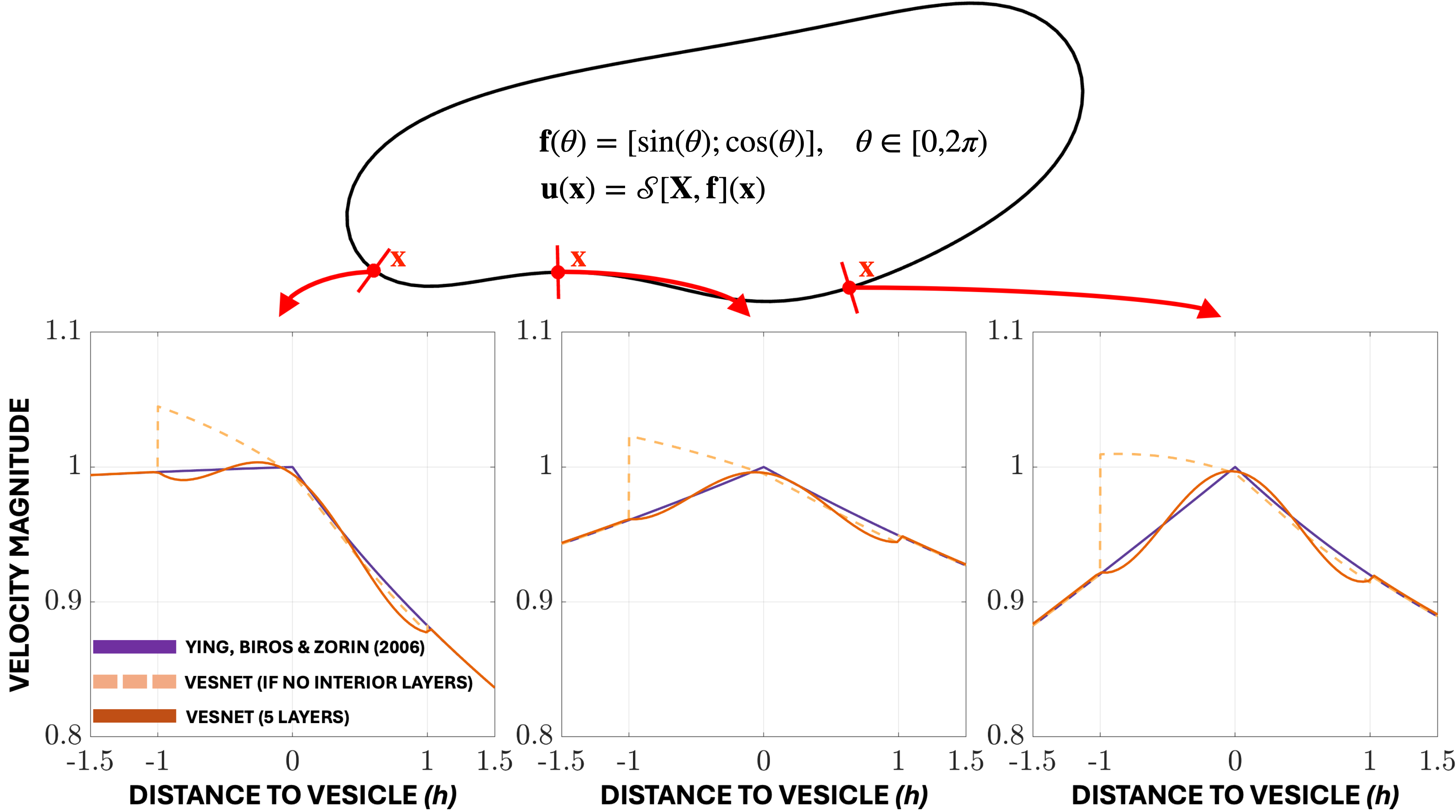}
    \caption{Evaluation of the proposed near-singular integration scheme with five layers at $(-h, -h/2, 0, h/2, h)$ normal distance to the vesicle for the near-field interactions. The proposed scheme is compared to its variation where the interior layers at $(-h, -h/2)$ are not used for the interpolation and to the near-singular integration scheme in \citet{ying-zorin-e06}. We choose three points normal to which the target points are created $\xx \in [-1.5h, 1.5h]$ (shown in red). We define a membrane force $\ff (\theta) = [\sin (\theta); \cos (\theta)]$ and evaluate the single layer integral to find the velocity at the target points. At the bottom, we show the velocity magnitude with respect to the distance to the vesicle. The proposed scheme uses the direct evaluation of the integral for $|d| > h$ and the interpolation of the velocity evaluated at the layers for $|d| \leq h$. Including the interior layers provides smooth interpolation across the vesicle surface, improving the continuity and stability of the reconstructed velocity field near the interface.
    }
    \label{FigNear}
\end{figure}

Here we show the difference between performing near-singular integration with the proposed five layers in \secref{sec:nearsing} and only three layers outside a vesicle. In \figref{FigNear} we pick a deformed vesicle $\XX$ and define a membrane force $\ff (\theta) = [\sin (\theta); \cos (\theta)]$ where $\theta = [0, 2\pi)$. Then, we evaluate the single layer integral \eqref{eq:single_layer} at target points $\xx$ which are distributed along three lines $[-1.5h, 1.5h]$ (shown in red) normal to three points on the vesicle (shown as red circle). We perform the integral evaluation using the algorithm in \citet{ying-zorin-e06} (purple lines), \vesnet with five layers around a vesicle at $(-h, -h/2, 0, h/2, h)$ (orange lines) and \vesnet with three layers at $(0, h/2, h)$. The plots show the magnitude of the single layer integral (i.e., velocity) normalized by the maximum velocity along these lines obtained using the algorithm in \citet{ying-zorin-e06} in the $y$-axis with respect to the distance to the vesicle. The direct integration (not shown here) goes to infinity as the target point approaches to the vesicle. The results show that the proposed scheme with fiver layers used in \vesnet can accurately capture the nearly singular integrals evaluated across the vesicle if the interior layers are also used for the interpolation. Otherwise, the scheme gives wrong velocity and leads to discontinuities in the velocity across the vesicle, which is accentuated when vesicles collide. 

\section{Error distribution of neural networks}\label{sec:appendixErrorsNN}
\begin{figure}
    \centering
    \includegraphics[width=0.5\textwidth]{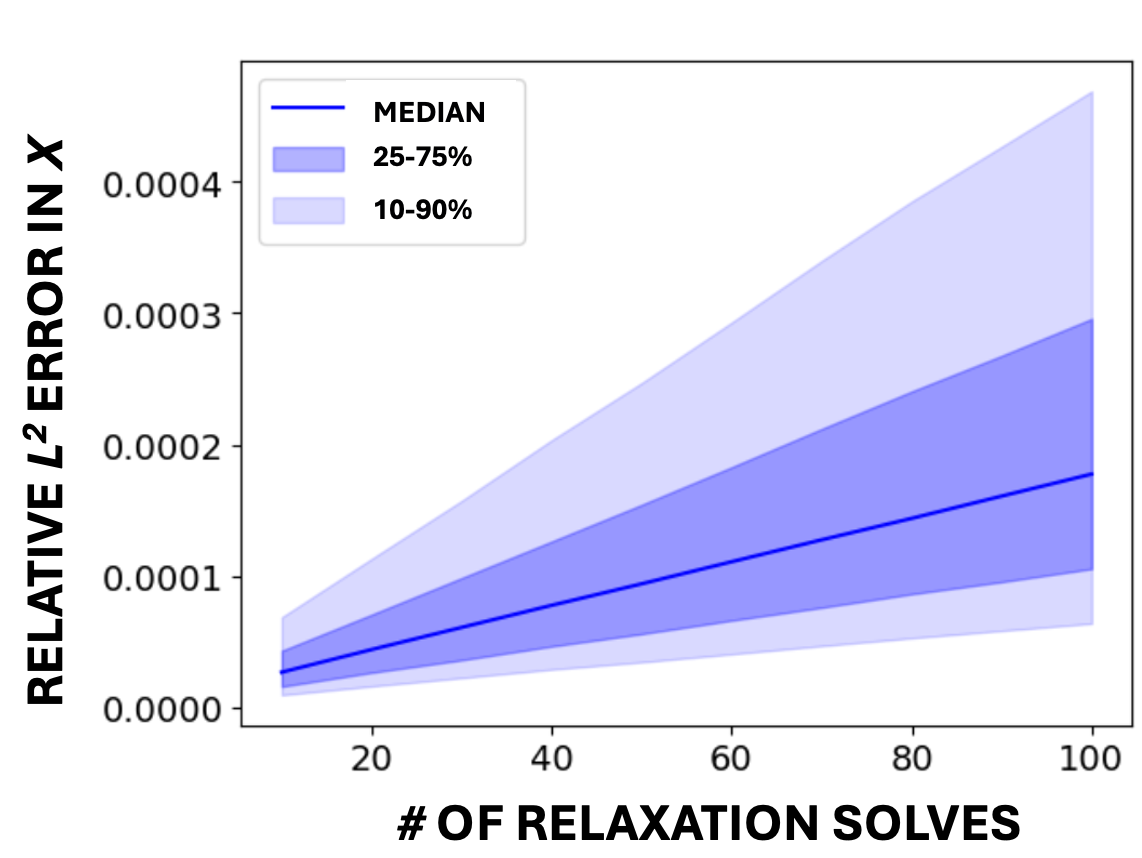}
    \caption{The evolution of relative error in $\XX$ of the relaxation network during 100 relaxation steps. The distribution of errors for  3200 vesicle samples is shown. Solid blue line shows the median error, while shaded regions indicate percentile spreads.}
    \label{FigErrRelax}
\end{figure}

\begin{figure}
    \centering
    \includegraphics[width=0.95\textwidth]{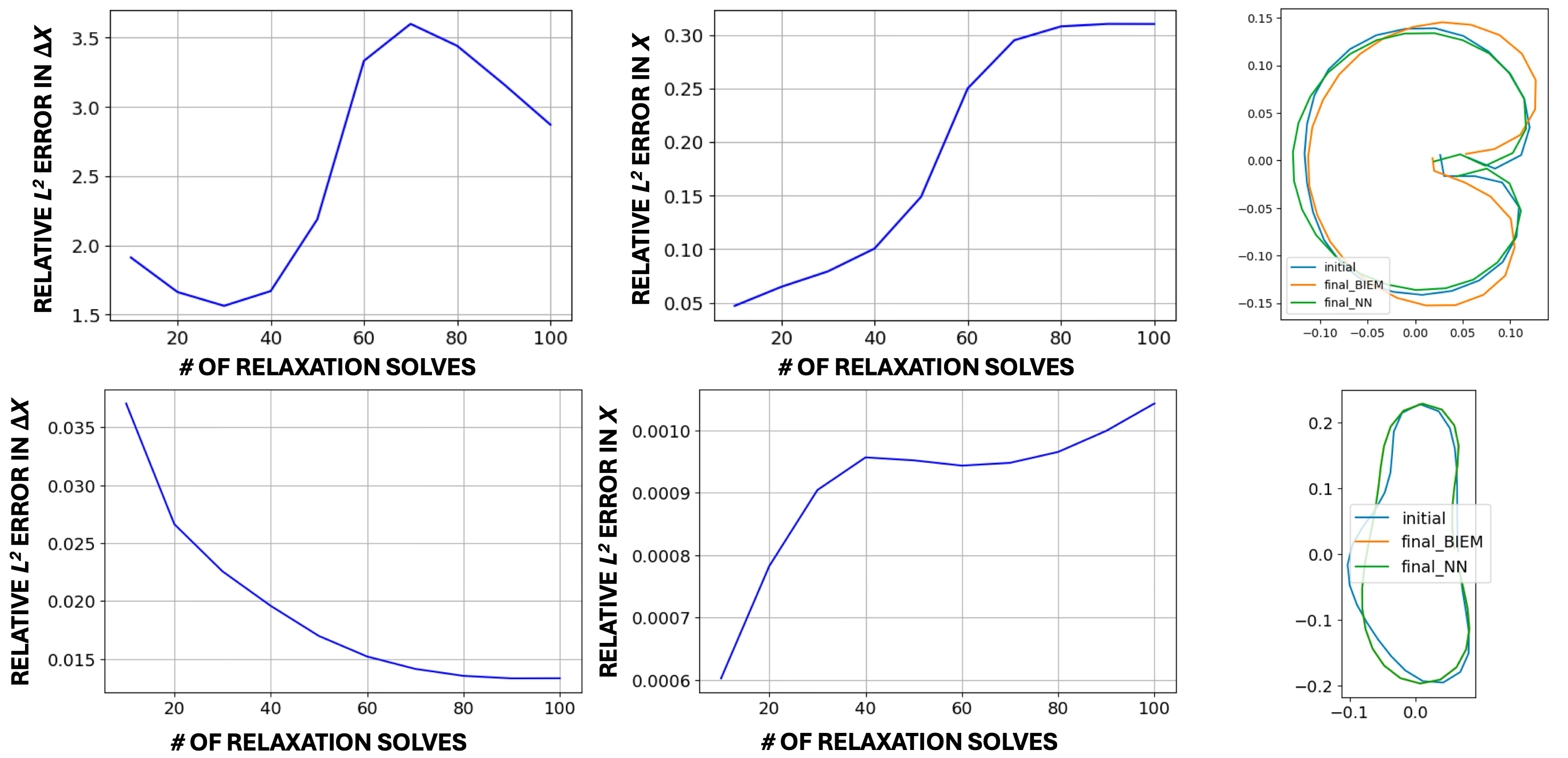}
    \caption{A large relaxation error example (upper row) and a normal error example (lower row). Evolutions of the relative error of displacement $\frac{\|\Delta \XX_{\text{NN}} - \Delta \XX_{\text{ground truth}}\|_2}{\|\Delta \XX_{\text{ground truth}}\|_2}$ (left column) as well as the relative error of shape $\frac{\|\XX_{\text{NN}} - \XX_{\text{ground truth}}\|_2}{\|\XX_{\text{ground truth}}\|_2}$  (middle column) are shown. The vesicle shapes are plotted in the right column, where we show initial shapes and final shapes computed by BIEM and \vesnet\ after 100 relaxation steps. The upper example exhibits large error due to the unusual shape configuration.}
    \label{FigRelaxExample}
\end{figure}

\begin{figure}
    \centering
    \includegraphics[width=0.5\textwidth]{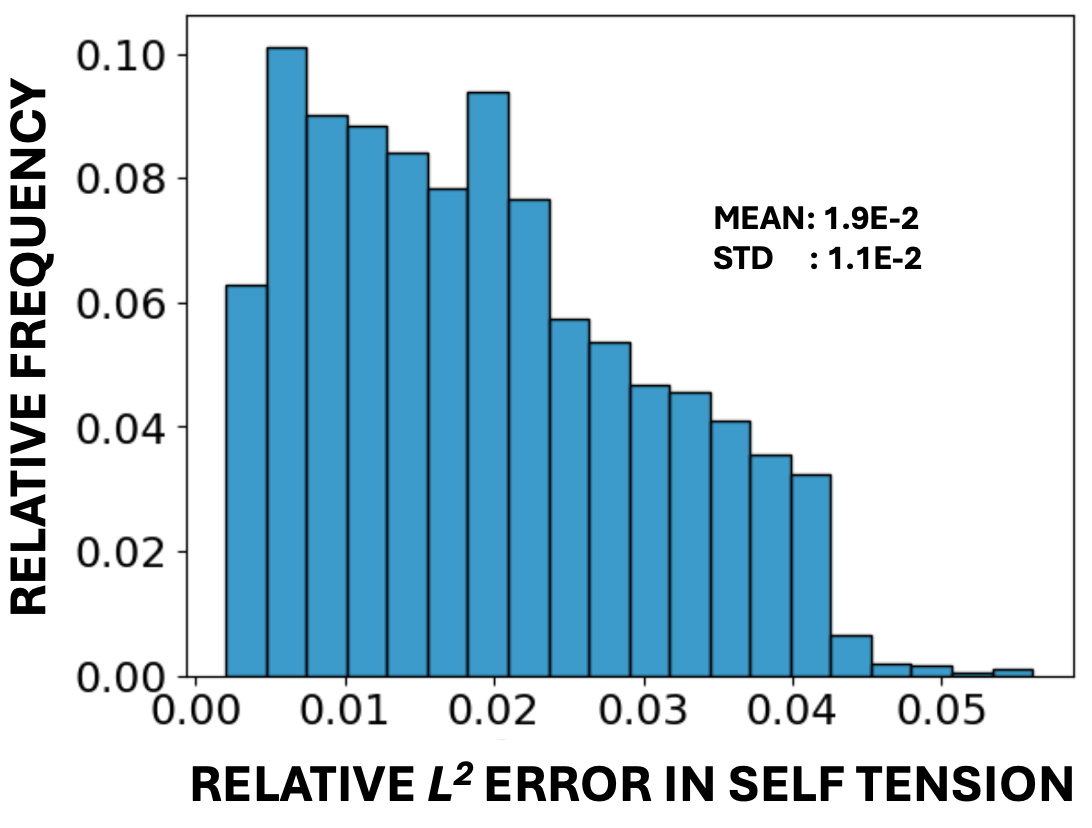}
    \caption{Error distribution of the self tension network.}
    \label{FigErrSelften}
\end{figure}

\begin{figure}
    \centering
    \includegraphics[width=0.95\textwidth]{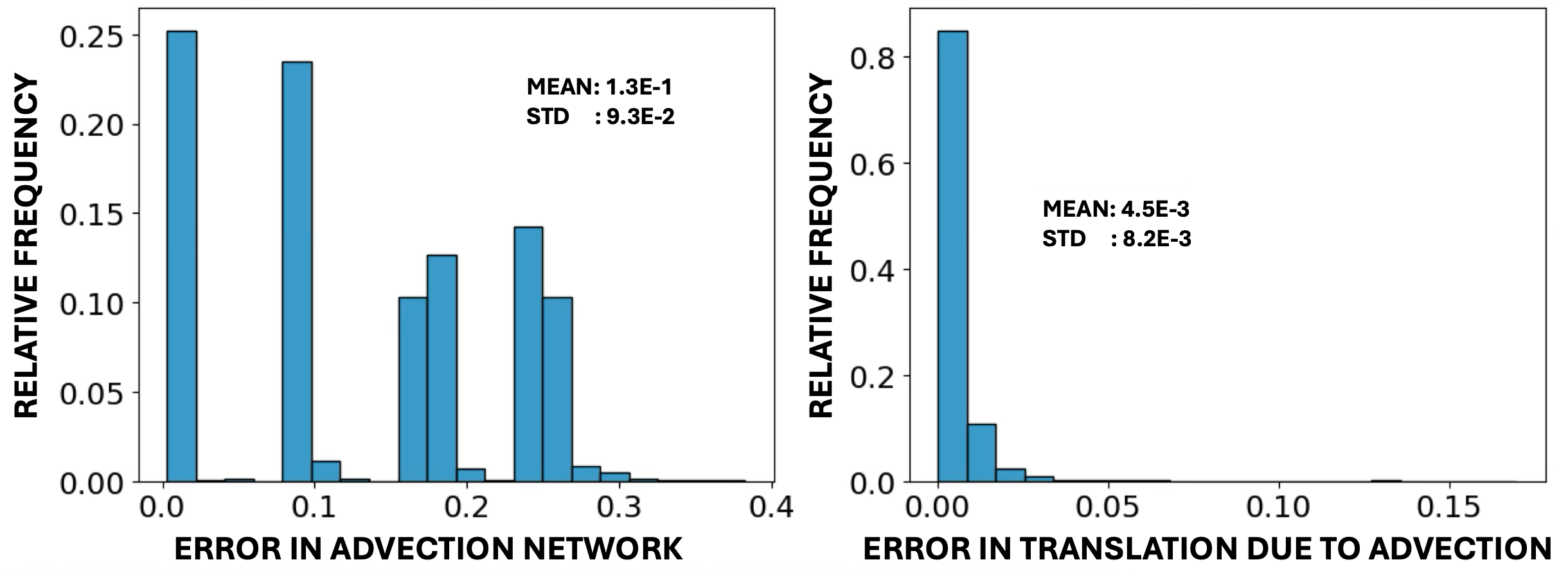}
    \caption{Error distributions of the advection network. The left figure shows the network prediction errors averaged over different Fourier modes, where errors of each mode are similarly distributed. The right figure shows the errors in the resulted advection.}
    \label{FigErrAdv}
\end{figure}

\begin{figure}
    \centering
    \includegraphics[width=0.95\textwidth]{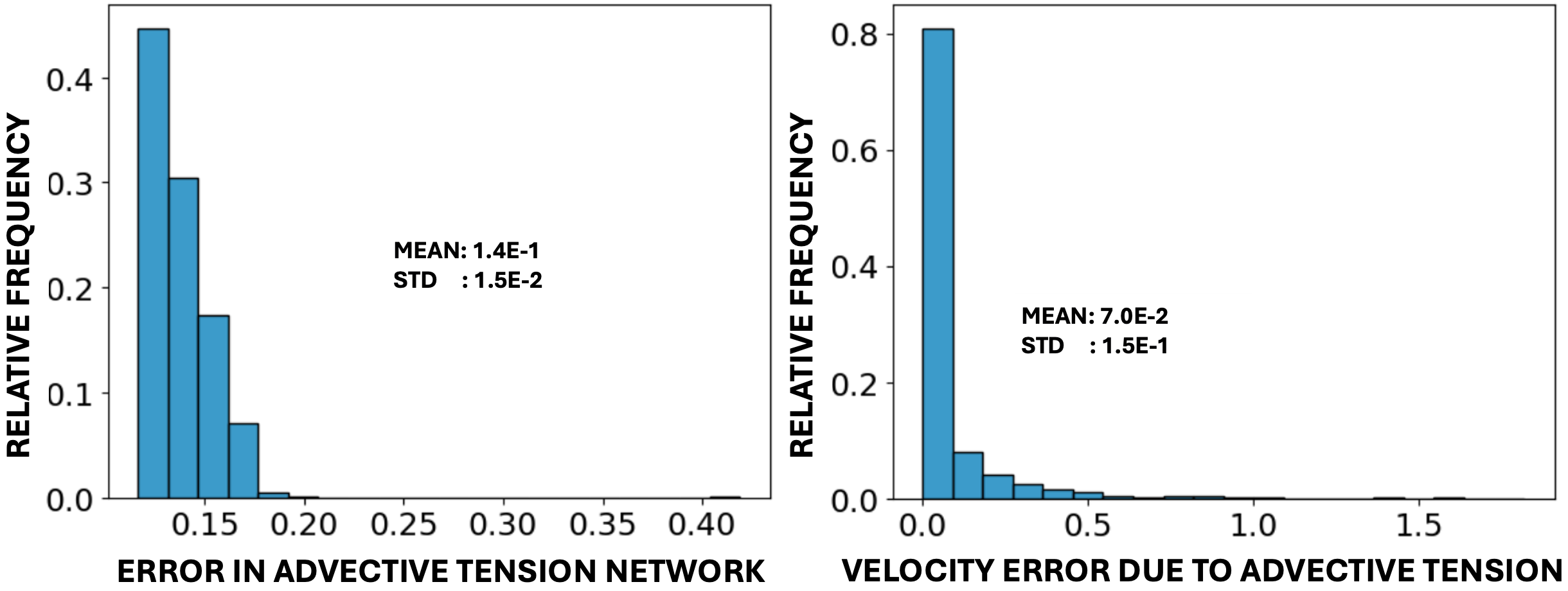}
    \caption{Error distributions of the advective tensions. The left figure shows the network prediction errors averaged over different Fourier modes, where errors of each mode are similarly distributed. The right figure shows the errors in the resulted velocity. Large errors in the right figure are due to small ground truth values.}
    \label{FigErrAdvten}
\end{figure}

\begin{figure}
    \centering
    \includegraphics[width=0.95\textwidth]{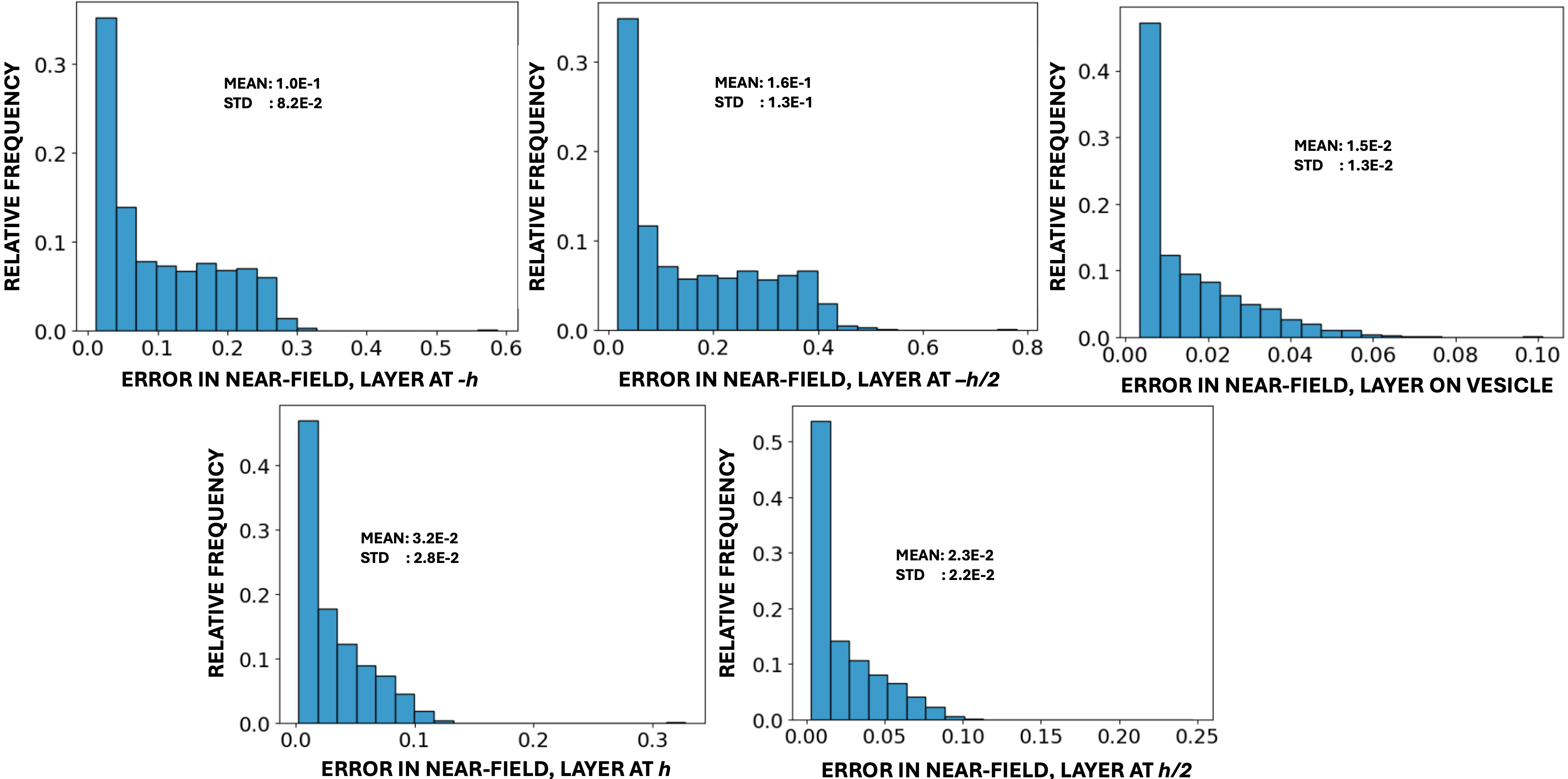}
    \caption{Error distributions of the near field networks, at different layers. Each figure shows the network prediction errors averaged over different Fourier modes. }
    \label{FigErrNear}
\end{figure}

Here we show the error distributions of each neural network component. The testing vesicles are drawn from 100 snapshots of a Taylor-Green flow simulation with BIEM. Every snapshot has 32 vesicles, leading to a total of 3200 samples. We report $L^2$ relative error, which is $\frac{\|\text{prediction} - \text{ground truth}\|_2}{\|\text{ground truth}\|_2}$. For relaxation, we study the stability of this operator by tracking the error in $\XX$ during 100 consecutive relaxation steps. We also show two vesicle examples in \figref{FigRelaxExample}. Note that these two examples are not samples from Taylor-Green flow simulation.


\bibliographystyle{jfm}
\bibliography{jfm}

\end{document}